\documentclass[acmsmall]{acmart}

\AtBeginDocument{%
  \providecommand\BibTeX{{%
    \normalfont B\kern-0.5em{\scshape i\kern-0.25em b}\kern-0.8em\TeX}}}

\usepackage{soul}
\usepackage{graphicx}
\usepackage{booktabs}
\usepackage{float}
\usepackage{overpic}
\usepackage{float}  
\usepackage{subfloat}
\usepackage{stfloats} 
\usepackage[skip=0.5\baselineskip]{caption}
\usepackage{bm}
\usepackage{amsmath}
\usepackage{booktabs}
\usepackage{multirow}
\usepackage{multicol}
\usepackage{enumitem}
\usepackage{subcaption}
\usepackage{threeparttable}
\usepackage{xspace}
\usepackage{color}
\usepackage[normalem]{ulem}
\usepackage{algorithmicx,algorithm}
\usepackage{algpseudocode}
\usepackage{algorithm,algpseudocode}

\usepackage{xspace}
\newcommand{\etal}{\emph{et al.}\xspace}
\newcommand{\eg}{\emph{e.g.,}\xspace}
\newcommand{\ie}{\emph{i.e.,}\xspace}
\newcommand{\etc}{\emph{etc.}\xspace}

\setcopyright{acmcopyright}
\copyrightyear{2024}
\acmYear{2024}
\acmDOI{XXXXXXX.XXXXXXX}

\begin{document}

\title{Multimodal Recommender Systems: A Survey}

\author{Qidong Liu}
\authornote{These authors contributed equally to this research.}
\email{liuqidong@stu.xjtu.edu.cn}
\orcid{0000-0002-0751-2602}
\affiliation{%
  \institution{Xi'an Jiaotong University \& City University of Hong Kong}
  \city{Xi'an}
  \country{China}
}

\author{Jiaxi Hu}
\email{jiaxihu2-c@my.cityu.edu.hk}
\authornotemark[1]
\orcid{0009-0008-3857-9069}
\affiliation{%
  \institution{City University of Hong Kong}
  \city{Hong Kong}
  \country{China}
  }

\author{Yutian Xiao}
\email{yutiaxiao2-c@my.cityu.edu.hk}
\authornotemark[1]
\orcid{0000-0002-8276-7920}
\affiliation{%
  \institution{City University of Hong Kong}
  \city{Hong Kong}
  \country{China}
}

\author{Xiangyu Zhao}
\authornote{Corresponding Author.}
\email{xianzhao@cityu.edu.hk}
\orcid{0000-0003-2926-4416}
\affiliation{%
  \institution{City University of Hong Kong}
  \city{Hong Kong}
  \country{China}
}

\author{Jingtong Gao}
\email{jt.g@my.cityu.edu.hk}
\orcid{0000-0002-4470-5972}
\affiliation{%
  \institution{City University of Hong Kong}
  \city{Hong Kong}
  \country{China}
}

\author{Wanyu Wang}
\email{wanyuwang4-c@my.cityu.edu.hk}
\orcid{0000-0001-5976-0707}
\affiliation{%
  \institution{City University of Hong Kong}
  \city{Hong Kong}
  \country{China}
}

\author{Qing Li}
\email{qing-prof.li@polyu.edu.hk}
\orcid{0000-0003-3370-471X}
\affiliation{%
  \institution{The Hong Kong Polytechnic University}
  \city{Hong Kong}
  \country{China}
}

\author{Jiliang Tang}
\email{tangjili@msu.edu}
\orcid{0000-0001-7125-3898}
\affiliation{%
  \institution{Michigan State University}
  \city{East Lansing}
  \country{USA}
}

\renewcommand{\shortauthors}{Qidong Liu, Jiaxi Hu, Yutian Xiao \etal}

\begin{abstract}
  The recommender system (RS) has been an integral toolkit of online services. They are equipped with various deep learning techniques to model user preference based on identifier and attribute information. With the emergence of multimedia services, such as short videos, news and \etc, understanding these contents while recommending becomes critical. Besides, multimodal features are also helpful in alleviating the problem of data sparsity in RS. Thus, \textbf{M}ultimodal \textbf{R}ecommender \textbf{S}ystem (MRS) has attracted much attention from both academia and industry recently. In this paper, we will give a comprehensive survey of the MRS models, mainly from technical views. First, we conclude the general procedures and major challenges for MRS. Then, we introduce the existing MRS models according to four categories, \ie \textbf{Modality Encoder}, \textbf{Feature Interaction}, \textbf{Feature Enhancement} and \textbf{Model Optimization}. Besides, to make it convenient for those who want to research this field, we also summarize the dataset and code resources. Finally, we discuss some promising future directions of MRS and conclude this paper.
  To access more details of the surveyed papers, such as implementation code, we open source a repository\footnote{https://github.com/Applied-Machine-Learning-Lab/Awesome-Multimodal-Recommender-Systems}.

\end{abstract}

\ccsdesc[500]{Information systems~Recommender systems}
\ccsdesc[500]{Information systems~Multimedia and multimodal retrieval}

\keywords{Recommender Systems; Multi-Modal; Multi-Media}


\maketitle

\section{Introduction}
With the advancement of the internet, many multimedia online services are emerging, such as fashion recommendation~\cite{deldjoo2022review} and \etc These multimedia applications give a chance to push the RS towards the path of understanding recommended items, which is much beneficial. On the one hand, understanding can help RS make use of abundant multimodal information of items to alleviate the problems of data sparsity~\cite{deldjoo2022review}. On the other hand, it assists the RS in knowing about the user's preference more deeply from a semantic level. Considering the prevalence of multimedia services, multimodal recommender system (MRS) is promising to become a general pattern of RS in the future. Therefore, more research has focused on MRS recently, and a review to survey and categorize them is in urgent need.

\begin{figure}[t]
\centering
\includegraphics[width=0.7\linewidth]{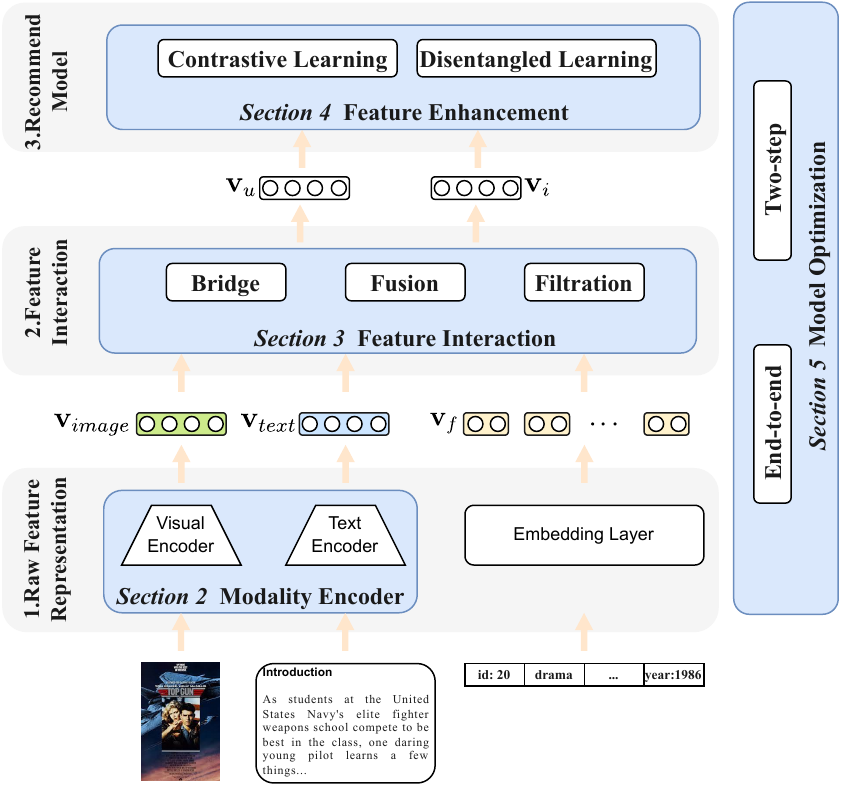}
\caption{The general procedures of multimodal recommender system.}
\label{fig_procedure}
\end{figure}

In general, the recommender system focuses on collaborative or side information, which refers to the identifier (abbreviated to id) and tabular features of items, such as genera and published year. By comparison, in an MRS, multimodal features, such as image, audio and text, play a vital role. For simplicity, we define the MRS as: \textit{the recommender system for the items with multimodal features}. In the following subsections, we will introduce the general procedures and our taxonomy to make the survey more readable.

\subsection{Procedures of MRS}
Based on the input items of MRS, we conclude the unified procedures for MRS, as Figure~\ref{fig_procedure} shows. There are three procedures: \textbf{Raw Feature Representation}, \textbf{Feature Interaction} and \textbf{Recommendation}. We take the movie recommendation as an example to illustrate the general procedures as follows: 

\noindent \textbf{Raw Feature Representation}. Each movie possesses two types of features: tabular features that describe its important characteristics using numerical values or classifications (such as genera or year), and multimodal features that depict the movie across various modalities of representation (such as poster image and textual introduction). To handle the tabular features, general recommender systems often adopt an embedding layer to transform sparse discrete features into dense vectors~\cite{javed2021review}. 
Specifically, the embedding layer treats each feature field as a discrete set and maps it to a fixed-length representation.
However, multimodal features often have varying formats and cannot share such an embedding layer. Therefore, the multimodal features are often fed into different modality encoders to extract comprehensive representations. The modality encoders are often general architectures used in other fields, such as ViT~\cite{dosovitskiy2020image} for images and Bert~\cite{devlin2018bert} for texts. Then, we can get the representations of tabular features and multimodal features (\ie image and text) for each item, denoted as $\mathbf{v}_f, \mathbf{v}_{image}$ and $\mathbf{v}_{text}$.

\noindent \textbf{Feature Interaction}. We get the representations of different modalities for each item, but they are in different semantic spaces. Besides, different users also have various preferences for modalities~\cite{wei2019mmgcn}. Therefore, in this procedure, MRS seeks to fuse and interact multimodal representations $\mathbf{v}_f, \mathbf{v}_{image}$ and $\mathbf{v}_{text}$ to get a unified item and user representations, which are often used to get the recommendation list~\cite{han2022vlsnr,liu2019user}.

\noindent \textbf{Recommendation}. After the second procedure, we get the representations of user and item, denoted as $\mathbf{v}_u$ and $\mathbf{v}_i$. 
The general recommendation models absorb these two representations and give the recommendation probabilities for different items~\cite{koren2009matrix}. However, the problem of data sparsity always degrades recommendation performance. Therefore, many research studies~\cite{liu2022disentangled,liu2022contrastive} propose to enhance the representations by incorporating multimodal information.

\subsection{Taxonomy}

\noindent Multimodal features bring the chance to alleviate the problem of data sparsity. However, due to the complexity and heterogeneity of the multimodal features, there are some challenges for each procedure and the whole process of the MRS. Such challenges block the benefit of multimodal features and even have adverse effects on RS. Therefore, to make the best of multimodal information, existing research focuses on facing one or some of these challenges. Then, the major challenges and corresponding solutions are listed according to the procedure of MRS.
\begin{itemize}[leftmargin=*]
    \item \textit{\textbf{Challenge 1}: For the raw multimodal inputs, \eg image or texts, how to get representations from the complex modality features}. The embedding layer has been widely adopted in general RS to learn the representation from raw features. 
    Nevertheless, it is hard to get representations from complex images or texts. 
    For example, extracting useful information from the raw poster or introduction of movies is difficult but vital evidently.
    Thanks to the advancements in computer vision and natural language processing, many encoders can be borrowed by MRS to get the representations, such as ViT~\cite{dosovitskiy2020image} for images. In this paper, we denote all encoders that handle multimodal features as \textbf{Modality Encoder} (in Section~\ref{sec_encoder}).

    \item \textit{\textbf{Challenge 2}: For the feature interaction procedure, how to fuse the modality features in different semantic spaces and get various preferences for each modality}. The heterogeneous nature of multimodal features causes difficulty in learning item representation and user preference for MRS. 
    Case in point, the features of the movie poster and introduction are in totally distinct representation space, which hinders the imputation of the user's preference.
    To face this challenge, many works design the model by extracting the relationships between users, items, and modalities. These works are categorized into \textbf{Feature Interaction} technique (in Section~\ref{sec_interaction}). 

    \item \textit{\textbf{Challenge 3}: For the recommendation procedure, how to get comprehensive representations for recommendation models under the data-sparse condition}. Though the multimodal features enrich the information of items, the sparsity problem still exists because of the small volume of interaction records in RS.
    As an example, the sparsity of one typical movie recommendation dataset, Movielens-1M\footnote{https://grouplens.org/datasets/movielens/}, is beyond $95\%$.
    Compared with general RS, multimodal features can be utilized further to enhance the representation of the user and item. We denote this line of works as \textbf{Feature Enhancement} (in Section~\ref{sec_enhancement}). 

    \item \textit{\textbf{Challenge 4:} For the whole process of MRS, how to optimize the lightweight recommendation models and parameterized modality encoder}.
    Taking a movie recommendation model~\cite{liang2023mmmlp} as the example, the parameter scale of its text encoder for movie introduction is about $110$M, while the basic RS accounts for less than $10$M.
    To solve the optimization problem, some MRS works propose novel techniques, which are clustered into \textbf{Model Optimization} (in Section~\ref{sec_opt}).

\end{itemize}

Based on the four challenges mentioned above, we organize the rest of this paper according to the corresponding technical solutions, \ie \textbf{Modality Encoder}, \textbf{Feature Interaction}, \textbf{Feature Enhancement} and \textbf{Model Optimization}. As far as we know, this survey is totally different from the existing two MRS surveys. One review~\cite{deldjoo2020recommender} organized the research following the different modalities in real applications. The other latest survey~\cite{zhou2023comprehensive} paid more attention to the RS itself while ignoring the characteristics of MRS. By comparison, our survey organizes the description concerning various types of techniques, especially for multimodal, which may help readers better understand the general MRS architecture. Also, we try to collect all recent works to help readers know about the recent advancements in this field.

\begin{table}[t]
\centering
\caption{Category for Modality Encoder}
\resizebox{1\columnwidth}{!}{
\begin{tabular}{c|c|l}
\toprule[1.5pt]
\textbf{Modality} & \textbf{Category} & \textbf{Related Works} \\
\midrule
\midrule
\multirow{3}{*}{Visual Encoder} 
& CNN & \cite{chen2019personalized}, \cite{liu2019nrpa}, \cite{kim2022mario}, \cite{zhang2020multi} \cite{mu2022learning}, \cite{wang2021multimodal}, \cite{chen2021cmbf}, \cite{chen2019pog}, \cite{liu2022disentangled}, \cite{liu2022implicit}, \cite{lin2019explainable}, \cite{han2022modality}, \cite{liu2022multi}, \cite{liu2022megcf}, \cite{liu2023semantic}, \cite{zhou2023attention}, \cite{li2023multimodal} \\
& ResNet & \cite{liu2019user}, \cite{pan2022multimodal}, \cite{ni2022two}, \cite{liu2021pre}, \cite{wang2022multimodal}, \cite{sun2020multi}, \cite{wu2021mm}, \cite{wei2019mmgcn}, \cite{wang2021dualgnn}, \cite{hou2019explainable}, \cite{hu2023adaptive}, \cite{bian2023multi} \\
 & Transformer & \cite{chen2022hybrid}, \cite{han2022vlsnr} \\ 
 \midrule
\multirow{5}{*}{Textual Encoder} & Word2vec & \cite{liu2019nrpa}, \cite{cao2022cross}, \cite{sun2020multi}, \cite{wang2021multimodal}, \cite{wei2019mmgcn}, \cite{wang2021dualgnn}, \cite{han2022modality} \\
 & RNN & \cite{xu2020recommendation}, \cite{lin2019explainable} \\
 & CNN & \cite{wang2020enhanced}, \cite{chen2019pog} \\
 & Sentence-transformer & \cite{kim2022mario}, \cite{zhang2022latent}, \cite{zhang2021mining}, \cite{mu2022learning}, \cite{zhou2022tale}, \cite{zhou2023attention}, \cite{li2023multimodal}, \cite{wei2023multi}, \cite{wei2024promptmm}  \\
 & Bert & \cite{liu2019user}, \cite{pan2022multimodal}, \cite{liu2021noninvasive}, \cite{ni2022two}, \cite{liu2021pre}, \cite{wang2022multimodal}, \cite{chen2022hybrid}, \cite{wu2021mm}, \cite{lv2019interest}, \cite{liu2022disentangled}, \cite{han2022vlsnr}, \cite{hu2023adaptive}, \cite{bian2023multi}, \cite{liu2023semantic}, \cite{liang2023mmmlp} \\ 
 \midrule
\multirow{1}{*}{Other Modality Encoder} & Published Feature & \cite{tao2020mgat}, \cite{zhang2022latent}, \cite{zhang2021mining}, \cite{yinwei2021grcn}, \cite{wei2019mmgcn}, \cite{wang2021dualgnn}, \cite{yi2022multi}, \cite{yu2023multi}, \cite{shang2023enhancing}, \cite{zhou2023bootstrap}, \cite{wei2023multi}, \cite{wei2024promptmm} \\ 
\bottomrule[1.5pt]
\end{tabular}
}
\label{tab:encoder}
\end{table}

\section{Modality Encoder} \label{sec_encoder}

The multimodal features of items are critical in constructing more specific user interests and enhancing model interpretability. In the context of recommendation tasks, item id information is typically represented by dense vectors obtained through a trainable embedding table. However, the multimodal features of items require corresponding encoders to obtain dense feature representations to extract more comprehensive information. In this section, in addition to pre-published feature vectors, we provide a brief introduction to commonly used encoders for three types of multimodal features, \ie \textbf{Visual}, \textbf{Textual}, and \textbf{Other modalities}, and provide detailed encoder information for existing MRS models in Table \ref{tab:encoder}.

\begin{itemize}[leftmargin=*]
    \item \textbf{Visual Encoder}: Visual feature is one vital modality in MRS, such as the poster for movie recommendation and clothing image for fashion recommendation. Most early MRS use a CNN-based pre-trained model as an image encoder. It compresses and extracts features through convolution and pooling operations from raw pixel information. For example, MMGCN \cite{wei2019mmgcn} adopts ResNet \cite{he2016deep} to extract information from the visual information of the news. POG \cite{chen2019personalized} uses VGG as an image encoder for cloth pictures. 
    Recently, visual pre-training models based on the transformer \cite{vaswani2017attention} architecture have achieved better performance, and some MRS models \cite{chen2022hybrid,han2022vlsnr} have started to use ViT \cite{dosovitskiy2020image} to extract visual features. So, we mainly categorize the visual encoder into CNN-based, ResNet-based and Transformer-based.
    \item \textbf{Textual Encoder}:
    Textual information, such as item descriptions, often contain more semantic information than images, making them more suitable for enhancing user interest modeling~\cite{zhou2023mmrec}. Some MRS models \cite{han2022modality, sun2020multi} typically  utilize GloVe \cite{pennington2014glove} or Word2Vec \cite{mikolov2013efficient} to extract text features. In addition, some models also use encoders such as Text-CNN \cite{kim-2014-convolutional}. With the development of natural language models, text encoders have gradually been standardized to Bert \cite{devlin2018bert}.  In conclusion, we categorize textual encoders into five lines, \ie Bert-based, RNN-based, CNN-based, Sentence-transformer-based and Word2vec-based.
    \item \textbf{Other Modality Encoder}: For acoustic and video data, they are often converted into text or visual information before input into the textual or visual encoders. Besides, some methods directly use the pre-published feature vectors contained in original datasets to model the acoustic and video modalities.
\end{itemize}

\begin{figure}[t]
    \centering
    \includegraphics[width=\textwidth]{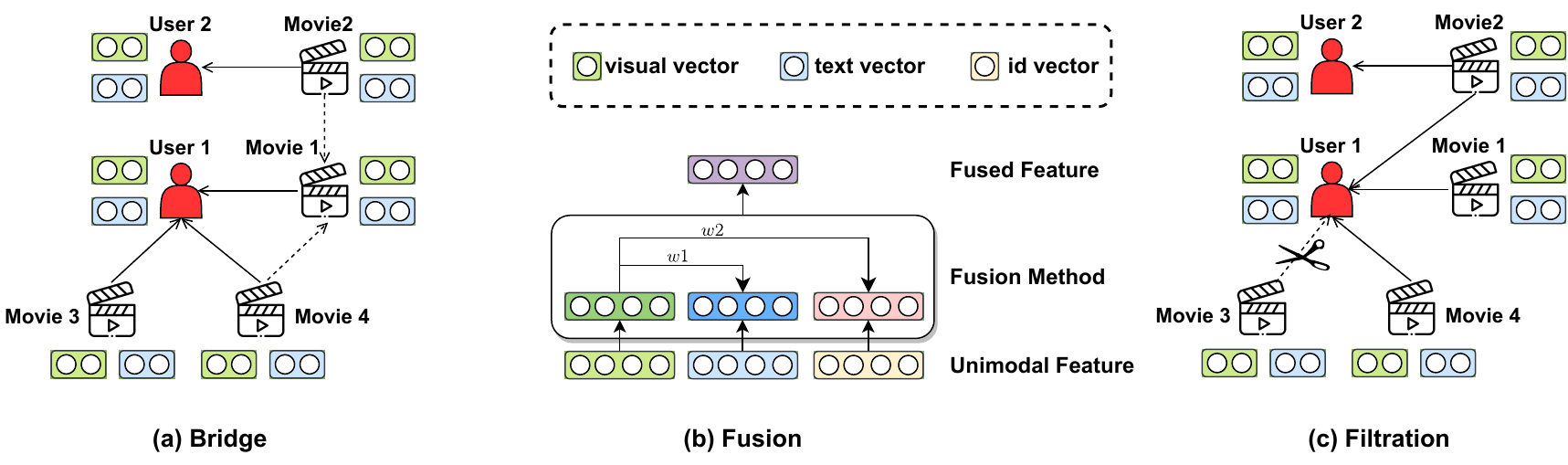}
    \caption{The illustration of three types of feature interaction.}
    \label{fig_interaction}
\end{figure}

\section{Feature Interaction}
\label{sec_interaction}
Multimodal data refers to various modalities of description information. 
Since they are sparse and in different semantic spaces, connecting them to the recommendation task is essential. The feature interaction can realize the nonlinear transformation of various feature spaces of different modalities into common space, finally elevating the performance and generalization of the recommendation model. As shown in Figure~\ref{fig_interaction}, we categorize interactions into three types: \textbf{Bridge}, \textbf{Fusion}, and \textbf{Filtration}. These three types of techniques implement interaction from various views, so they can be applied to one MRS model simultaneously. For readability, we also categorize the existing works based on their interaction type in Table~\ref{tab:interaction}.

\subsection{Bridge}
Here bridge refers to the construction of a multimodal information transfer channel. 
It focuses on capturing the inter-relationship between users and items considering multimodal information.
The difference between multimedia and traditional recommendation is that the items contain rich multimedia information. Most early works simply use multimodal content to enhance the item expression, but they often ignore the interactions between users and items. The message-passing mechanism of graph neural networks can enhance user representation through information exchange between users and items and further capture the user's preference for different modal information. Figure~\ref{fig_interaction}(a) gives an example: many models get user $1$ preference by aggregating interacted items for each modality. Besides, the modality representation of movie $1$ can be derived from the latent item-item graph.
In this subsection, we will introduce the methods for how to build bridges in MRS.

\subsubsection{\textbf{User-item Graph}}
Leveraging the information exchange between users and items, users' preferences for different modalities can be captured. Therefore, some works utilize the user-item graph.
MMGCN \cite{wei2019mmgcn} establishes a user-item bipartite graph for each modality. For each node, the topology of adjacent nodes and the modality information of the item can be used to update the feature expression of the node. Based on MMGCN, GRCN \cite{yinwei2021grcn} improves the performance of recommendations by adaptively modifying the graph's structure during model training to delete incorrect interaction data (users clicked uninterested videos). Although these methods have achieved great success in performance, these methods are still limited by using a unified way to fuse user preferences of different modalities, ignoring the difference in the degree of user preference for different modalities. 
In other words, giving equal weight to each modality may result in the sub-optimal performance of the model. 
To solve this problem, DualGNN \cite{wang2021dualgnn} utilizes the correlation between users to learn user preferences based on the bipartite and user co-occurrence graph. Also, MMGCL \cite{yi2022multi} designs a new multimodal graph contrastive learning method to solve this problem. MMGCL uses modal edge loss and modal masking to generate user-item graphs and introduces a novel negative sampling technique to learn the correlation between modalities. MGAT \cite{tao2020mgat} introduces an attention mechanism based on MMGCN, which is conducive to adaptively capturing user preferences for different modalities. Moreover, MGAT uses the gated attention mechanism to judge the user's preference for different modalities, which can capture relatively complex interaction patterns contained in user behaviors.

\subsubsection{\textbf{Item-item Graph}}
The above works focus on using multimodal features to model user-item interactions while ignoring latent semantic item-item structures. Reasonable use of item-item structures is conducive to better learning item representation and improving model performance. 
For instance, LATTICE \cite{zhang2021mining} constructs an item-item graph for each modality based on the user-item bipartite graph. It aggregates them to obtain the latent item graphs. MICRO \cite{zhang2022latent} also constructs an item-item graph for each modality. Unlike LATTICE, MICRO adopts a new comparison method to fuse features after performing graph convolution. However, these works do not take into account the differences in preferences between various specific user groups. 
Furthermore, HCGCN \cite{mu2022learning} proposes a clustering graph convolutional network, which first groups item-item and user-item graphs and then learns user preferences through dynamic graph clustering. Besides, inspired by the recent success of pre-training models, PMGT \cite{liu2021pre} proposes a pre-trained graph transformer referring to Bert's structure and provides a unified view of project relationships and their associated side information in a multimodal form. BGCN \cite{chang2020bundle}, as a model in bundle recommendation, unifies the user-item interaction, user-bundle interaction, and bundle-item affiliation into a heterogeneous graph, using graph convolution to extract fine-gained features. Cross-CBR \cite{ma2022crosscbr} builds the user-bundle graph, the user-item diagram, and the item-bundle graph, using contrastive learning to align them from the bundle and item views. 

\begin{table}[t]
\centering
\caption{Category for Feature Interaction}
\resizebox{1\columnwidth}{!}{
\begin{tabular}{c|c|c|l}
\toprule[1.5pt]
\textbf{Interaction} & \textbf{Goal} & \textbf{Category} & \textbf{Related Works} \\ 
\midrule
\midrule
\multirow{3}{*}{Bridge} & \multirow{3}{*}{\shortstack{Capture inter-relationship \\ between users and items}} 
& User-item Graph & \cite{tao2020mgat}, \cite{ni2022two}, \cite{wang2021dualgnn}, \cite{wei2019mmgcn}, \cite{yi2022multi}, \cite{yu2023multi}, \cite{liu2023semantic}, \cite{zhou2023bootstrap}, \cite{wei2023multi}, \cite{wei2024promptmm} \\
 & & Item-item Graph & \cite{zhang2022latent}, \cite{zhang2021mining}, \cite{mu2022learning}, \cite{liu2021pre}, \cite{chang2020bundle}, \cite{ma2022crosscbr}, \cite{zhang2022latent}, \cite{hu2023adaptive}, \cite{yu2023multi}, \cite{zhou2023attention} \\
 & & Knowledge Graph & \cite{wang2020enhanced}, \cite{wang2019multi}, \cite{cao2022cross}, \cite{wang2022multimodal}, \cite{chen2022hybrid}, \cite{sun2020multi}, \cite{liu2022multi} \\ 
 \midrule
\multirow{5}{*}{Fusion} & \multirow{5}{*}{\shortstack{Combine various preference \\ to modalities}} 
& Coarse-grained Attention & \cite{liu2022contrastive}, \cite{pan2022multimodal}, \cite{liu2019user}, \cite{chen2021cmbf}, \cite{yu2023multi}, \cite{zhou2023attention}, \cite{wei2023multi} \\
 & & \multirow{2}{*}{Fine-grained Attention} & \cite{chen2019personalized}, \cite{kim2022mario}, \cite{xiao2020deep}, \cite{lian2021multi}, \cite{tao2020mgat}, \cite{ni2022two}, \cite{liu2021pre}, \cite{chen2022hybrid}, \cite{han2022modality}, \cite{liu2022disentangled}, \cite{kim2022mario}, \\ & & & \cite{liu2022implicit}, \cite{lei2021understanding}, \cite{liu2022multi}, \cite{chen2019pog}, \cite{hou2019explainable},\cite{lin2019explainable}, \cite{zhu2022combo}, \cite{wu2021mm}, \cite{li2022miner} \\
 & & Combined Attention & \cite{liu2019nrpa}, \cite{liu2021noninvasive}, \cite{han2022vlsnr}, \cite{hu2023adaptive} \\
 & & Other Fusion Methods & \cite{wang2021multimodal}, \cite{chen2021curriculum}, \cite{lv2019interest}, \cite{lv2019interest}, \cite{zhang2020multi}, \cite{xu2020recommendation}, \cite{bian2023multi}, \cite{zhou2023bootstrap}, \cite{liang2023mmmlp} \\ 
 \midrule
Filtration & Filter out noisy data & Filtration & \cite{sun2020multi}, \cite{zhou2022tale}, \cite{yinwei2021grcn}, \cite{liu2022megcf}, \cite{yi2021multi}, \cite{shang2023enhancing}, \cite{li2023multimodal}, \cite{zhong2024mirror} \\ 
\bottomrule[1.5pt]
\end{tabular}
}
\label{tab:interaction}
\end{table}

\subsubsection{\textbf{Knowledge Graph}}
Knowledge graphs (KG) are widely used because they can provide auxiliary information for recommender systems. To combine the KG and MRS, many researchers introduce each modality of items to KG as an entity.
MKGAT \cite{sun2020multi} is the first model to introduce a knowledge graph into the multimodal recommendation. MKGAT proposes a multimodal graph attention technique to model multimodal knowledge graph from two aspects of entity information aggregation and entity relationship reasoning, respectively. Furthermore, a novel graph attention network is adopted to aggregate neighboring entities while considering the relations in the knowledge graph. SI-MKR \cite{wang2020enhanced} proposes an enhanced multimodal recommendation method based on alternate training and the knowledge graph representation based on MKR \cite{wang2019multi}. Besides, most multimodal recommender systems ignore the problem of data type diversity. SI-MKR solves it by adding user and item attribute information from the knowledge graph. By comparison, MMKGV \cite{liu2022multi} adopts a graph attention network for information dissemination and information aggregation on a knowledge graph, which combines multimodal information and uses the triplet reasoning relationship of the knowledge graph. CMCKG~\cite{cao2022cross} treats information from descriptive attributes and structural connections as two modals and learns node representation by maximizing consistency between these two views.

\subsection{Fusion}
In the multimodal recommendation scenario, the types and quantities of multimodal information of users and items are very large. So, it is necessary to fuse the different multimodal information to generate the feature vector for the recommendation task~\cite{li2023imf}.
Compared with bridge, fusion is more concerned about the multimodal intra-relationships of items. To be specific, it aims to combine various preferences into modalities. Since the inter- and intra-relationships are vital to learning comprehensive representations, many MRS models~\cite{tao2020mgat,ni2022two} even adopt both fusion and bridge. 
The attention mechanism is the most widely used feature fusion method, which can flexibly fuse multimodal information with different weights and focus. 
In this subsection, as shown in Figure~\ref{fig_interaction}(b), we first divide attention mechanisms by fusion granularity and then introduce some of the other fusion approaches that exist in the MRS.

\subsubsection{\textbf{Coarse-grained Attention}} 
Some models apply attention mechanisms to fuse information from multiple modalities at a coarse-grained level. For example, UVCAN~\cite{liu2019user} divides multimodal information into user-side and item-side, including their respective id information and side information. It uses multimodal data on the user side to generate fusion weights for the item side through self-attention. 
In addition to the user and item sides, some models merge information from different modal aspects. 
CMBF~\cite{chen2021cmbf} introduces the cross-attention mechanism to co-learn the semantic information of image and text modality, respectively, and then concatenate them.
Besides, the proportions of various modals are also different in some models. MML~\cite{pan2022multimodal} designs an attention layer based on id information and is assisted by visual and text information. Liu~\etal~\cite{liu2022contrastive} point out that each modal occupies the same position, and the self-attention mechanism determines the fusion weight. By comparison, HCGCN~\cite{mu2022learning} pays more attention to the visual and text information of the item itself.
Besides, MGCN~\cite{yu2023multi} proposes behavior-aware attention to combining the modalities with distinct importance. Zhou~\etal~\cite{zhou2023attention} highlights the difference between identity and other modalities and then designs multi-level attention for fusion.

\subsubsection{\textbf{Fine-gained Attention}} 
The multimodal data contains both global and fine-grained features, such as the tone of the audio recording or the pattern of clothing. Since coarse-grained fusion is often invasive and irreversible~\cite{liu2021noninvasive}, it will damage the original modality information and degrade the recommendation performance. Therefore, some works consider fine-grained fusion, which selectively fuses fine-grained information between different modalities. 

Fine-grained fusion is significant in the fashion recommendation scenario.
POG~\cite{chen2019pog} is a sizeable online clothing RS based on transformer architecture. The encoder excavates the deep features belonging to the collocation scheme in fashion images through multi-layer attention, which continuously realizes fine-grained integration. Compared with POG, NOR~\cite{lin2019explainable} applies both encoder-decoder transformer architecture, which contains fine-grained self-attention structures. It can generate the corresponding scheme description according to collocation information. 
Besides, to increase interpretability, EFRM~\cite{hou2019explainable} also designs a Semantic Extraction Network (SEN) to extract the local features, and finally fuses the two features with fine-grained attention preference. VECF~\cite{chen2019personalized} performs image segmentation to integrate image features of each patch with other modalities. 
UVCAN~\cite{liu2022implicit} conducts image segmentation of video screenshots like VECF and fuse image patches with id information and text information through the attention mechanism, respectively. MM-Rec~\cite{wu2021mm} first extracts the region of interest from the image of news through the target detection algorithm Mask-RCNN and then fuses POI with news content using co-attention. MINER \cite{li2022miner}, DMIN \cite{xiao2020deep}, SUM \cite{lian2021multi} all build interest representations of different aspects of the user by the multimodal information.

Some other models design unique internal structures for better fine-grained fusion. For instance, MKGformer~\cite{chen2022hybrid} achieves fine-grained fusion by sharing some QKV parameters and a related perceptual fusion module. MGAT~\cite{tao2020mgat} uses a gated attention mechanism to focus on the user's local preferences. MARIO~\cite{kim2022mario} predicts user preferences by considering the individual impact of each modality on each interaction. So, it designs a modality-aware attention mechanism to identify the influence of various modalities on each interaction and conducts point multiplication for different modalities.

\subsubsection{\textbf{Combined Attention}} Based on fine-grained fusion, some models design combined fusion structures, hoping that the fusion of fine-grained features can also preserve the aggregation of global information.
NOVA~\cite{liu2021noninvasive} introduces side information to the sequential recommendation. It points out that directly fusing different modal features with vanilla attention usually brings little effect or even degrades performance. So, it proposes a non-invasive attention mechanism with two branches, id embedding in separate ones to preserve interactive information in the fusion process. NRPA~\cite{liu2019nrpa} offers a personalized attention network, which considers user preferences represented by user comments. It uses personalized word-level attention to select more important words in comments for each user/item, and passes the comment information to the attention layer through fine-grained and coarse-grained fusion in turn. VLSNR~\cite{han2022vlsnr} is another application of sequential recommendation, \ie news recommendation. It can model users' temporary and long-term interests and realize fine-grained and coarse-grained fusion by multi-head attention and GRU network. 
Moreover, MMSR~\cite{hu2023adaptive} employs dual attention to preserve modalities' temporal order for the sequential recommendation.

\subsubsection{\textbf{Other Fusion Methods}}
In addition to fusing the multimodal information through attention weights, some works apply other simple methods, including concat operations~\cite{zhang2020multi}, and gating mechanism~\cite{liu2021noninvasive}. Nevertheless, they rarely appear alone and often in combination with the graph and attention mechanisms, as mentioned above. Existing work~\cite{liu2021noninvasive} has shown that simple interactions, if appropriately used, will not damage the recommendation effect, and can reduce the complexity of the model. Besides, some early models adopt structures such as RNN~\cite{han2022vlsnr}, attempting to model user temporal preferences through multimodal information.  
The other models fuse the multimodal feature through linear and nonlinear layers. Lv et.al.~\cite{lv2019interest} set a linear layer at the place to fuse the textual and visual features.
In MMT-Net~\cite{krishnan2020transfer}, three context invariants of restaurant data are artificially marked, and interaction is carried out through MLPs. 
Recently, more fabricated architectures have been developed for fusion, such as the mixture of expert~\cite{bian2023multi} and MLP mixer~\cite{liang2023mmmlp}.

\subsection{Filtration}
As multimodal data differs from user interaction data, it contains much information unrelated to user preferences. For example, as shown in Figure~\ref{fig_interaction}(c), the interaction between movie $3$ and user $1$ is noisy, which should be removed. Filtering out noisy data in multimodal recommendation tasks can usually improve the recommendation performance. It is worth noting that noise can exist in the interaction graph or multimodal feature itself, so filtration can be embedded in the bridge and fusion, respectively.

Some models use image processing to denoise. For example, VECF ~\cite{chen2019personalized} and UVCAN ~\cite{liu2022implicit} perform image segmentation to remove noise from the image so that they can better model the user's personalized interests. MM-Rec~\cite{wu2021mm} uses a target detection algorithm to select the significant margin of the image.

In addition, many structures based on graph neural networks are also used for denoising. Due to the sparsity of user-item interactions and the noise of item features, the representation of users and items learned through graph aggregation inherently contain noise. 
FREEDOM \cite{zhou2022tale} designs a degree-sensitive edge pruning method to denoise the user-item interaction graph.
GRCN \cite{yinwei2021grcn} detects whether the user accidentally interacts with a noisy item.
Unlike the GCN model, GRCN can adaptively adjust the structure of the interaction graph during training to identify and prune wrong interaction information. PMGCRN \cite{jia2022preference} also takes user interactions with uninterested items into account, but unlike GRCN, it handles mismatched interactions with an active attention mechanism to correct users' wrong preferences.
Besides, MEGCF \cite{liu2022megcf} focuses on the mismatch problem between multimodal feature extraction and user interest modeling. It firstly constructs a multimodal user-item graph and then uses sentiment information from comment data to fine-grained weight neighbor aggregation in the GCN module to filter noise. 
MAGAE~\cite{yi2021multi} is a model designed to handle uncertainty in multi-modal data. 
Besides, Shang~\etal~\cite{shang2023enhancing} firstly find out the problem of modality imbalance and propose to filter out the items with sensitive modalities. Then, MCLN~\cite{li2023multimodal} integrates a novel counterfactual framework to eliminate the noise.

\section{Feature Enhancement} \label{sec_enhancement}

\begin{figure}[t]
    \centering
    \includegraphics[width=\textwidth]{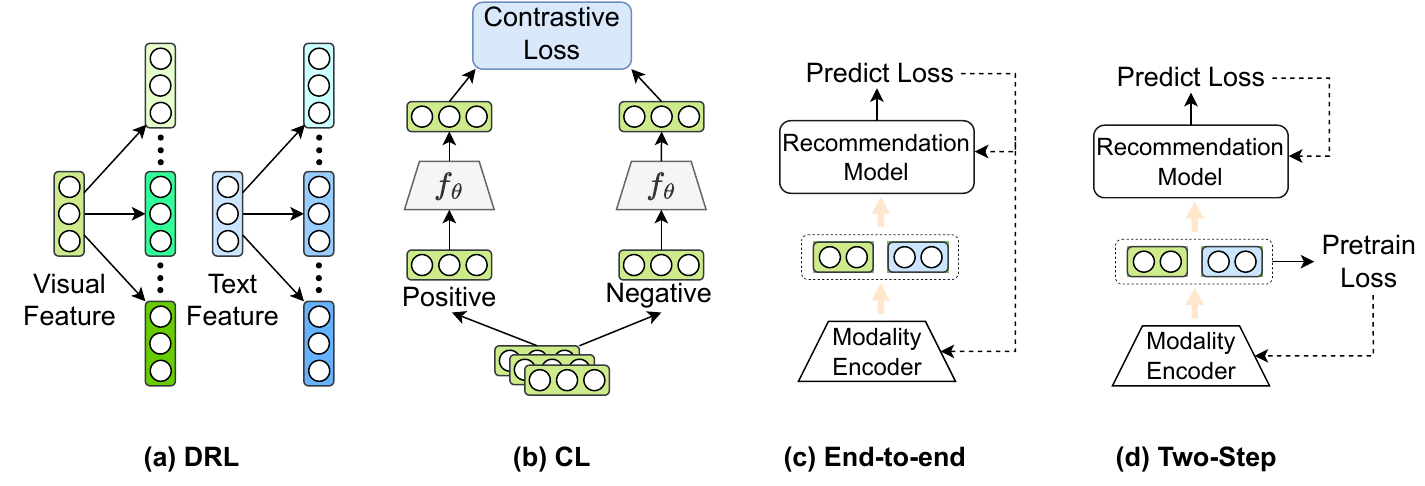}
    \caption{The illustration of feature enhancement and model optimization.}
    \label{fig:drl_en}
\end{figure}

Different modality representations of the same object have unique or common semantic information. 
Therefore, the recommendation performance and generalization of MRS can be significantly improved if the unique and common characteristics can be distinguished.
Recently, to solve this problem, some models are equipped with \textbf{Disentangled Representation Learning (DRL)} and \textbf{Contrastive Learning (CL)} to carry out feature enhancement based on interaction, as shown in Figure~\ref{fig:drl_en}(a) and (b).

\subsection{Disentangled Representation Learning}
The features of different modalities have various importance to the user's preference on a particular factor of the target item in RS. However, the representations of different factors in each modality are often entangled, so many researchers have introduced decomposition learning techniques to dig out the meticulous factors in user preference, such as DICER \cite {zhang2020content}, MacridVAE \cite{ma2019learning}, CDR \cite{chen2021curriculum}.
Besides, the multimodal recommendation is to discover helpful information formed by various hidden factors from multimodal data, which are highly entangled in complex ways. 
MDR \cite{wang2021multimodal} proposes a multimodal disentangled recommendation that can learn well-disentangled representations carrying complementary and standard information from different modalities. DMRL \cite{liu2022disentangled} considers the different contributions of various modality features for each disentanglement factor to capture user preferences. Furthermore, PAMD \cite{han2022modality} designs a disentangled encoder to extract modality-common features while preserving modality-specific features automatically. Besides, the designed contrastive learning guarantees the consistency and gap between separated modal representations. Compared with MacridVAE, SEM-MacridVAE~\cite{9720218} considers item semantic information when learning disentangled representations from user behaviors.

\subsection{Contrastive Learning}
Unlike DRL, contrastive learning methods enhance the representation by data augmentation, which is also helpful in handling the sparsity problem. 
Besides, Many works have introduced CL loss functions mainly for modality alignment. 

Liu~\etal~\cite{liu2022contrastive} propose a novel CL loss, which makes the different modal representations of the same item have semantic similarity. In addition, GHMFC~\cite{wang2022multimodal} constructs two contrastive learning modules, based on the entity embedding representations derived from the graph neural network. The two CL loss functions are in two directions, \ie text to image and image to text. Cross-CBR \cite{ma2022crosscbr} proposes a CL loss to align the graph representation from the bundle view and item view. MICRO~\cite{zhang2022latent} focuses on both shared modal information and specific modal information. In CMCKG~\cite{cao2022cross}, entity embeddings are obtained from both descriptive attributes and structural link information through knowledge graphs for contrastive loss. In HCGCN~\cite{mu2022learning}, to enforce visual and textual item features mapped into the same semantic space, it refers to CLIP~\cite{radford2021learning} that adopts contrastive learning and maximizes the similarity of correct visual-textual pair in a batch. 
Furthermore, Zhou~\etal~\cite{zhou2023bootstrap} and Wei~\etal~\cite{wei2023multi} both design the modality-based contrastive loss for alignment and digging out the inter-relationships between distinct modalities.

Because the core of contrastive learning is to mine the relationship between positive and negative samples, many models adopt data augmentation methods to construct positive samples in recommendation scenarios. MGMC~\cite{zhao2021multimodal} designs a graph enhancement to augment the samples and introduces meta-learning to increase model generalization. MML~\cite{pan2022multimodal} is a sequential recommendation model that expands the training data by constructing a subset of the user's historical purchase item sequence. LHBPMR~\cite{mu2022learning} selects items with similar preferences from the graph convolution to construct positive samples. MMGCL~\cite{yi2022multi} constructs positive samples by modal edge dropping and modal masking. Also, Victor~\cite{lei2021understanding} firstly constructs samples through Chinese semantics. Combo-Fashion \cite{zhu2022combo} is a bundle fashion recommendation model, so it constructs negative and positive fashion matching schemes. Most of the existing models consider removing information that does not belong to user preferences in multimodal data. 
By comparison, UMPR~\cite{xu2020recommendation} directly constructs a loss that describes the difference between visual positive and negative samples.

\section{Model Optimization} \label{sec_opt}

Unlike traditional recommendation tasks, due to the existence of multimodal information, the computational requirements for model training are greatly increased when multimodal encoders and RS are trained together. Therefore, the MRS can be divided into two categories during training: \textbf{End-to-end training} and \textbf{Two-step training}. As shown in Figure~\ref{fig:drl_en}(c), End-to-end training can update the parameters of all layers in the model with each gradient obtained through backpropagation. By comparison, the two-step training includes the first stage of pretraining multimodal encoders and the second stage of task-oriented optimization, which is illustrated in Figure~\ref{fig:drl_en}(d).

\subsection{End-to-end Training}
Since multimodal recommender systems use pictures, texts, audio and other multimedia information, some common encoders in other fields, such as Vit ~\cite{dosovitskiy2020image}, Resnet~\cite{he2016deep}, Bert~\cite{devlin2018bert}, are often adopted when processing these multimodal data. The parameters of these pretrained models are often very huge. For example, the number of parameters of Vit-Base ~\cite{dosovitskiy2020image} reaches 86M, which is a great challenge for computing resources. 
To solve this problem, most MRS adopt pretrained encoders directly and only train the recommendation model in an end-to-end pattern. NOVA~\cite{liu2021noninvasive} and VLSNR~\cite{han2022vlsnr} use a pretrained encoder to encode images and text features, then embed the resulting multimodal feature vectors through the model and recommends for users. They show that introducing multimodal data without updating encoder parameters can also improve the recommendation performance. Liu~\etal~\cite{liu2022contrastive} propose to fine-tune the encoder's parameters with only 100 epochs by recommendation and contrastive loss. 
In particular, MG~\cite{zhong2024mirror} eliminates the noise contained in multimodal inputs by a gradient strategy.
Recently, some researchers~\cite {wang2022transrec,zhang2023multimodal} only utilize multimodal features of items by pretrained encoders, which can free MRS from the limitation of item identity in a specific dataset.

Some end-to-end methods also aim to reduce the amount of computation while improving the recommendation performance. They often decrease the number of parameters required to be updated when training. For instance, MKGformer~\cite{chen2022hybrid} is a multi-layer transformer structure where many attention layer parameters are shared to reduce computation. FREEDOM ~\cite{zhou2022tale} is designed to freeze some parameters of graph structure, dramatically reducing memory costs, and achieving a denoising effect.

\subsection{Two-step Training}

Compared with the end-to-end pattern, the two-stage training scheme can target downstream tasks better, but it requests much higher computing resources. Thus, few MRS adopt two-step training. 
PMGT~\cite{liu2021pre} proposes a pretrained graph transformer referring to Bert's structure. It learns item representations with two objectives: graph structure reconstruction and masked node feature reconstruction. In POG~\cite{chen2019pog}, it pretrains a transformer to learn the fashion matching knowledge, and then recommends for users through a cloth generation model. Besides, it is common in sequential recommendation, where it is difficult to train the model in an end-to-end scheme. For example, in the pretraining stage, MML~\cite{pan2022multimodal} first trains the meta-learner through meta-learning to increase model generalization, then trains the item embedding generator in the second stage. Besides, TESM~\cite{ni2022two} and Victor~\cite{lei2021understanding} pretrain a well-designed graph neural network and a video transformer, respectively. Recently, some more advanced techniques have been adapted for higher training efficiency, such as knowledge distillation and prompt tuning. As for the former one, SGFD~\cite{liu2023semantic} distills a lighter modality encoder from a pretrained modality encoder, when finetuning for the recommendation task. Also, PromptMM~\cite{wei2024promptmm} proposes a pretrain-prompt scheme to achieve easier finetuning and higher task adaptability.

\section{Applications and Resources} \label{sec:application}

Nowadays, when users browse the online shopping platform, they will receive a large amount of multimodal information about items, which will influence users' behavior imperceptibly. Though most researchers aim to propose a general MRS model that can be applied to all applications, it is better to design a unique model for some typical applications. For example, in fashion recommendation applications, users are often tempted to buy clothing because of the style of the clothing that they do not need. POG~\cite{chen2019pog} proposes to utilize the image and title of the clothing item to predict whether the style is compatible and preferred by the user.
Besides, the content of items is important for News recommendation, so the textual feature is highlighted in the related researches~\cite{han2022vlsnr,li2022miner}.
Normally, the general RS can also be adapted to these applications, however, they show markedly inferior performance compared with MRS models~\cite{liang2023mmmlp,chen2019pog,zhu2022combo,li2022miner}.

Dataset is one necessary resource to research MRS, especially for these typical applications.  
Therefore, to ease access to such vital resources, we summarize several popular datasets for MRS according to its applications in Table~\ref{table:statistics}.
This will guide the researchers to obtain these MRS datasets conveniently. Anyone who wishes to use these datasets can refer to the corresponding citations and websites for more details. 
In terms of the evaluation metrics for MRS models, they are often the same as the general RS, such as hit rate and normalized discounted cumulative gain.

\begin{table*}[!t]
        \centering
	\caption{Summary of the MRS datasets.}
        \setlength{\tabcolsep}{3pt}
	\label{table:statistics}
        \resizebox{1\columnwidth}{!}{
        \begin{threeparttable}
	\begin{tabular}{lcccccl}
		\toprule[1.5pt]
		Data & Field & Modality & Scale & Link\\ 
        \midrule
        \midrule
		\ Tiktok  & Micro-video & V,T,M,A  & 726K+  & \footnotesize https://paperswithcode.com/dataset/tiktok-dataset \\
		\ Kwai  & Micro-video &  V,T,M  &1 million+   & \footnotesize https://zenodo.org/record/4023390\#.Y9YZ6XZBw7c\\
        \midrule
		\ Movielens + IMDB & Movie & V,T  & 100k$\sim$25m & \footnotesize https://grouplens.org/datasets/movielens/ \\
        \ Douban  & Movie,Book,Music & V,T  & 1 million+ & \footnotesize https://github.com/FengZhu-Joey/GA-DTCDR/tree/main/Data \\
        \midrule
        \ Yelp  & POI & V,T,POI & 1 million+  & \footnotesize  https://www.yelp.com/dataset\\
        \ Amazon  & E-commerce & V,T & 100 million+  & \footnotesize https://cseweb.ucsd.edu/~jmcauley/datasets.html\#amazon\_reviews \\
        \midrule
        \ Book-Crossings  & Book & V,T & 1 million+  & \footnotesize http://www2.informatik.uni-freiburg.de/~cziegler/BX/  \\
        \ Amazon Books  & Book & V,T  & 3 million  & \footnotesize https://jmcauley.ucsd.edu/data/amazon/ \\
        \midrule
        \ Amazon Fashion  & Fashion & V,T  & 1 million  & \footnotesize https://jmcauley.ucsd.edu/data/amazon/ \\
        \ POG  & Fashion & V,T  & 1 million+ & \footnotesize https://drive.google.com/drive/folders/1xFdx5xuNXHGsUVG2VIohFTXf9S7G5veq \\
        \ Tianmao  & Fashion & V,T & 8 million+  & \footnotesize https://tianchi.aliyun.com/dataset/43 \\
        \ Taobao  & Fashion & V,T & 1 million+  & \footnotesize https://tianchi.aliyun.com/dataset/52 \\
        \midrule
        \ Tianchi News  & News & T  & 3 million+  & \footnotesize https://tianchi.aliyun.com/competition/entrance/531842/introduction\\
        \ MIND  & News & V,T  & 15 million+  & \footnotesize https://msnews.github.io/\\
        \midrule
        \ Last.FM  & Music & V,T,A  & 186 k+  & \footnotesize https://www.heywhale.com/mw/dataset/5cfe0526e727f8002c36b9d9/content \\
        \ MSD  & Music & T,A  & 48 million+  & \footnotesize http://millionsongdataset.com/challenge/ \\
        \bottomrule[1.5pt]
	\end{tabular} 
\begin{tablenotes}
\item[1] `V', `T', `M', `A' indicate the visual data, textual data, video data and acoustic data, respectively. 
\end{tablenotes}           
\end{threeparttable}
}
\end{table*}

As mentioned before, MRS often consists of several architecture components, which causes technical difficulty in implementation for practical systems. For example, the modality encoders with extensive parameters are difficult to deploy. Furthermore, it is hard to devise a unique training pipeline for MRS models because many branches exist, such as different interaction modules.
For data pre-processing, the settings of data split and filtration vary, which leads to challenges in reproduction.
To face these issues, two open-source benchmarks are helpful:

\begin{itemize}[leftmargin=*]
    \item \textbf{MMRec\footnote{https://github.com/enoche/MMRec}}: MMRec is a multimodal recommendation toolbox based on PyTorch. It integrates more than ten outstanding multimodal recommendation system models, such as MMGCN~\cite{wei2019mmgcn}.
    \item \textbf{Cornac\footnote{https://github.com/PreferredAI/cornac}}~\cite{salah2020cornac}: Cornac is a comparative framework for multimodal recommender systems. It derives the whole experimental procedures for MRS, \ie data, models, metrics and experiment. Besides, cornac is highly compatible with mainstream deep learning frameworks such as TensorFlow and PyTorch.
\end{itemize}

\section{Challenges and Future Directions}

To inspire the researchers who want to devote themselves to this field, we list several existing challenges for promising research:

\begin{itemize}[leftmargin=*]

    \item \textbf{A Universal Solution.}
    It is worth noting that though some methods for different stages in a model are proposed~\cite{lei2021understanding}, there is no up-to-date universal solution with the combinations of these techniques provided.
    
    \item \textbf{Model Interpretability}. The complexity of multimodal models can make it difficult to understand and interpret the recommendations generated by the system, which can limit the trust and transparency of the system. Though few pioneers~\cite{hou2019explainable,chen2019personalized} refer to it, it still needs to be explored.

    \item \textbf{Computational Complexity}. MRS requires large amounts of computational resources due to the parameter-intensive modality encoder, making it challenging to train the models on large datasets and populations. Also, the complexity of multimodal data and models can increase the inference cost and time required for recommendation generation, making it challenging for real-time applications.

    \item \textbf{Risk of Overfitting}.
    Due to the sparsity of the MRS data and informative representation obtained from the fabricated modality encoders, the MRS models are inclined to suffer from overfitting.

    \item \textbf{Privacy}.
    Though multimodal information can benefit recommender systems by alleviating data sparsity, it also increases the risk of privacy leakage. How to protect individual privacy under the condition of affluent multimodal information is also a great challenge for the researchers.

    \item \textbf{Large General MRS Dataset}.
    Currently, the scale of the MRS dataset is still limited, and the modalities covered are not extensive enough. In addition, the quality and availability of data for different modalities may vary, which can affect the accuracy and reliability of the recommendations.
    Therefore, a large high-quality MRS dataset with abundant modalities is urgently needed.

    \item \textbf{Incomplete and Biased Data}.
    In real-world applications, the multimodal data is often incomplete or biased. For example, one specific modality may be missed during the data collection. Besides, the practical interactions are often skewed by popularity. Addressing these two data challenges will accelerate the applications of MRS to industrial.

\end{itemize} 

\noindent Besides, we also give out several future directions as follows:

\begin{itemize}[leftmargin=*]
    \item \textbf{Cross-modal Representation Learning}.
    Developing better methods for cross-modal representation learning, such as transfer learning and large-scale pre-training encoder is expected to lead to more effective and efficient MRS.

    \item \textbf{Integration with Other Technologies}. An integration with technologies such as augmented reality and virtual reality is expected to enhance user experience and provide new opportunities for multimodal recommendation.

    \item \textbf{Utilization of Multimodal Large Language Models}. The multimodal large language models (MLLM) have shown brilliant understanding and reasoning abilities. Thus, the adaptation of MLLM to MRS is relatively promising~\cite{wang2023large}.
\end{itemize}

\section{Conclusion}

MRS is becoming one of the leading directions in RS, benefitting from its aggregation advantage on different modalities. In this paper, we introduce taxonomies of MRS, \ie modality encoder, feature interaction, feature enhancement and model optimization based on challenges faced in different modeling stages. We also summarize the dataset and open-source codes. 
At last, some challenges and future directions of MRS are proposed to inspire further research.

\begin{acks}
    This research was partially supported by Research Impact Fund (No.R1015-23), APRC - CityU New Research Initiatives (No.9610565, Start-up Grant for New Faculty of CityU), CityU - HKIDS Early Career Research Grant (No.9360163), Hong Kong ITC Innovation and Technology Fund Midstream Research Programme for Universities Project (No.ITS/034/22MS), Hong Kong Environmental and Conservation Fund (No. 88/2022), and SIRG - CityU Strategic Interdisciplinary Research Grant (No.7020046), Huawei (Huawei Innovation Research Program), Tencent (CCF-Tencent Open Fund, Tencent Rhino-Bird Focused Research Program), Ant Group (CCF-Ant Research Fund, Ant Group Research Fund), Alibaba (CCF-Alimama Tech Kangaroo Fund (No. 2024002)), CCF-BaiChuan-Ebtech Foundation Model Fund, and Kuaishou.
\end{acks}

\bibliographystyle{ACM-Reference-Format}
\bibliography{main}


\begin{thebibliography}{85}


\ifx \showCODEN    \undefined \def \showCODEN     #1{\unskip}     \fi
\ifx \showDOI      \undefined \def \showDOI       #1{#1}\fi
\ifx \showISBNx    \undefined \def \showISBNx     #1{\unskip}     \fi
\ifx \showISBNxiii \undefined \def \showISBNxiii  #1{\unskip}     \fi
\ifx \showISSN     \undefined \def \showISSN      #1{\unskip}     \fi
\ifx \showLCCN     \undefined \def \showLCCN      #1{\unskip}     \fi
\ifx \shownote     \undefined \def \shownote      #1{#1}          \fi
\ifx \showarticletitle \undefined \def \showarticletitle #1{#1}   \fi
\ifx \showURL      \undefined \def \showURL       {\relax}        \fi
\providecommand\bibfield[2]{#2}
\providecommand\bibinfo[2]{#2}
\providecommand\natexlab[1]{#1}
\providecommand\showeprint[2][]{arXiv:#2}

\bibitem[Bian et~al\mbox{.}(2023)]%
        {bian2023multi}
\bibfield{author}{\bibinfo{person}{Shuqing Bian}, \bibinfo{person}{Xingyu Pan},
  \bibinfo{person}{Wayne~Xin Zhao}, \bibinfo{person}{Jinpeng Wang},
  \bibinfo{person}{Chuyuan Wang}, {and} \bibinfo{person}{Ji-Rong Wen}.}
  \bibinfo{year}{2023}\natexlab{}.
\newblock \showarticletitle{Multi-modal mixture of experts represetation
  learning for sequential recommendation}. In \bibinfo{booktitle}{\emph{Proc.
  of CIKM}}.
\newblock


\bibitem[Cao et~al\mbox{.}(2022)]%
        {cao2022cross}
\bibfield{author}{\bibinfo{person}{Xianshuai Cao}, \bibinfo{person}{Yuliang
  Shi}, \bibinfo{person}{Jihu Wang}, \bibinfo{person}{Han Yu},
  \bibinfo{person}{Xinjun Wang}, {and} \bibinfo{person}{Zhongmin Yan}.}
  \bibinfo{year}{2022}\natexlab{}.
\newblock \showarticletitle{Cross-modal Knowledge Graph Contrastive Learning
  for Machine Learning Method Recommendation}. In
  \bibinfo{booktitle}{\emph{Proc. of ACM MM}}.
\newblock


\bibitem[Chang et~al\mbox{.}(2020)]%
        {chang2020bundle}
\bibfield{author}{\bibinfo{person}{Jianxin Chang}, \bibinfo{person}{Chen Gao},
  \bibinfo{person}{Xiangnan He}, \bibinfo{person}{Depeng Jin}, {and}
  \bibinfo{person}{Yong Li}.} \bibinfo{year}{2020}\natexlab{}.
\newblock \showarticletitle{Bundle recommendation with graph convolutional
  networks}. In \bibinfo{booktitle}{\emph{Proc. of SIGIR}}.
\newblock


\bibitem[Chen et~al\mbox{.}(2021a)]%
        {chen2021curriculum}
\bibfield{author}{\bibinfo{person}{Hong Chen}, \bibinfo{person}{Yudong Chen},
  \bibinfo{person}{Xin Wang}, \bibinfo{person}{Ruobing Xie},
  \bibinfo{person}{Rui Wang}, \bibinfo{person}{Feng Xia}, {and}
  \bibinfo{person}{Wenwu Zhu}.} \bibinfo{year}{2021}\natexlab{a}.
\newblock \showarticletitle{Curriculum Disentangled Recommendation with Noisy
  Multi-feedback}.
\newblock \bibinfo{journal}{\emph{Proc. of NeurIPS}} (\bibinfo{year}{2021}).
\newblock


\bibitem[Chen et~al\mbox{.}(2019b)]%
        {chen2019pog}
\bibfield{author}{\bibinfo{person}{Wen Chen}, \bibinfo{person}{Pipei Huang},
  \bibinfo{person}{Jiaming Xu}, \bibinfo{person}{Xin Guo},
  \bibinfo{person}{Cheng Guo}, \bibinfo{person}{Fei Sun}, \bibinfo{person}{Chao
  Li}, \bibinfo{person}{Andreas Pfadler}, \bibinfo{person}{Huan Zhao}, {and}
  \bibinfo{person}{Binqiang Zhao}.} \bibinfo{year}{2019}\natexlab{b}.
\newblock \showarticletitle{POG: personalized outfit generation for fashion
  recommendation at Alibaba iFashion}. In \bibinfo{booktitle}{\emph{Proc. of
  KDD}}.
\newblock


\bibitem[Chen et~al\mbox{.}(2019a)]%
        {chen2019personalized}
\bibfield{author}{\bibinfo{person}{Xu Chen}, \bibinfo{person}{Hanxiong Chen},
  \bibinfo{person}{Hongteng Xu}, \bibinfo{person}{Yongfeng Zhang},
  \bibinfo{person}{Yixin Cao}, \bibinfo{person}{Zheng Qin}, {and}
  \bibinfo{person}{Hongyuan Zha}.} \bibinfo{year}{2019}\natexlab{a}.
\newblock \showarticletitle{Personalized fashion recommendation with visual
  explanations based on multimodal attention network: Towards visually
  explainable recommendation}. In \bibinfo{booktitle}{\emph{Proc. of SIGIR}}.
\newblock


\bibitem[Chen et~al\mbox{.}(2021b)]%
        {chen2021cmbf}
\bibfield{author}{\bibinfo{person}{Xi Chen}, \bibinfo{person}{Yangsiyi Lu},
  \bibinfo{person}{Yuehai Wang}, {and} \bibinfo{person}{Jianyi Yang}.}
  \bibinfo{year}{2021}\natexlab{b}.
\newblock \showarticletitle{CMBF: Cross-Modal-Based Fusion Recommendation
  Algorithm}.
\newblock \bibinfo{journal}{\emph{Sensors}} (\bibinfo{year}{2021}).
\newblock


\bibitem[Chen et~al\mbox{.}(2022)]%
        {chen2022hybrid}
\bibfield{author}{\bibinfo{person}{Xiang Chen}, \bibinfo{person}{Ningyu Zhang},
  \bibinfo{person}{Lei Li}, \bibinfo{person}{Shumin Deng},
  \bibinfo{person}{Chuanqi Tan}, \bibinfo{person}{Changliang Xu},
  \bibinfo{person}{Fei Huang}, \bibinfo{person}{Luo Si}, {and}
  \bibinfo{person}{Huajun Chen}.} \bibinfo{year}{2022}\natexlab{}.
\newblock \showarticletitle{Hybrid Transformer with Multi-level Fusion for
  Multimodal Knowledge Graph Completion}.
\newblock \bibinfo{journal}{\emph{arXiv preprint arXiv:2205.02357}}
  (\bibinfo{year}{2022}).
\newblock


\bibitem[Deldjoo et~al\mbox{.}(2022)]%
        {deldjoo2022review}
\bibfield{author}{\bibinfo{person}{Yashar Deldjoo}, \bibinfo{person}{Fatemeh
  Nazary}, \bibinfo{person}{Arnau Ramisa}, \bibinfo{person}{Julian Mcauley},
  \bibinfo{person}{Giovanni Pellegrini}, \bibinfo{person}{Alejandro Bellogin},
  {and} \bibinfo{person}{Tommaso Di~Noia}.} \bibinfo{year}{2022}\natexlab{}.
\newblock \showarticletitle{A review of modern fashion recommender systems}.
\newblock \bibinfo{journal}{\emph{arXiv preprint arXiv:2202.02757}}
  (\bibinfo{year}{2022}).
\newblock


\bibitem[Deldjoo et~al\mbox{.}(2020)]%
        {deldjoo2020recommender}
\bibfield{author}{\bibinfo{person}{Yashar Deldjoo}, \bibinfo{person}{Markus
  Schedl}, \bibinfo{person}{Paolo Cremonesi}, {and} \bibinfo{person}{Gabriella
  Pasi}.} \bibinfo{year}{2020}\natexlab{}.
\newblock \showarticletitle{Recommender systems leveraging multimedia content}.
\newblock \bibinfo{journal}{\emph{ACM Computing Surveys (CSUR)}}
  (\bibinfo{year}{2020}).
\newblock


\bibitem[Devlin et~al\mbox{.}(2018)]%
        {devlin2018bert}
\bibfield{author}{\bibinfo{person}{Jacob Devlin}, \bibinfo{person}{Ming-Wei
  Chang}, \bibinfo{person}{Kenton Lee}, {and} \bibinfo{person}{Kristina
  Toutanova}.} \bibinfo{year}{2018}\natexlab{}.
\newblock \showarticletitle{Bert: Pre-training of deep bidirectional
  transformers for language understanding}.
\newblock \bibinfo{journal}{\emph{arXiv preprint arXiv:1810.04805}}
  (\bibinfo{year}{2018}).
\newblock


\bibitem[Dosovitskiy et~al\mbox{.}(2020)]%
        {dosovitskiy2020image}
\bibfield{author}{\bibinfo{person}{Alexey Dosovitskiy}, \bibinfo{person}{Lucas
  Beyer}, \bibinfo{person}{Alexander Kolesnikov}, \bibinfo{person}{Dirk
  Weissenborn}, \bibinfo{person}{Xiaohua Zhai}, \bibinfo{person}{Thomas
  Unterthiner}, \bibinfo{person}{Mostafa Dehghani}, \bibinfo{person}{Matthias
  Minderer}, \bibinfo{person}{Georg Heigold}, \bibinfo{person}{Sylvain Gelly},
  {et~al\mbox{.}}} \bibinfo{year}{2020}\natexlab{}.
\newblock \showarticletitle{An image is worth 16x16 words: Transformers for
  image recognition at scale}.
\newblock \bibinfo{journal}{\emph{arXiv preprint arXiv:2010.11929}}
  (\bibinfo{year}{2020}).
\newblock


\bibitem[Han et~al\mbox{.}(2022a)]%
        {han2022vlsnr}
\bibfield{author}{\bibinfo{person}{Songhao Han}, \bibinfo{person}{Wei Huang},
  {and} \bibinfo{person}{Xiaotian Luan}.} \bibinfo{year}{2022}\natexlab{a}.
\newblock \showarticletitle{VLSNR: Vision-Linguistics Coordination Time
  Sequence-aware News Recommendation}.
\newblock \bibinfo{journal}{\emph{arXiv preprint arXiv:2210.02946}}
  (\bibinfo{year}{2022}).
\newblock


\bibitem[Han et~al\mbox{.}(2022b)]%
        {han2022modality}
\bibfield{author}{\bibinfo{person}{Tengyue Han}, \bibinfo{person}{Pengfei
  Wang}, \bibinfo{person}{Shaozhang Niu}, {and} \bibinfo{person}{Chenliang
  Li}.} \bibinfo{year}{2022}\natexlab{b}.
\newblock \showarticletitle{Modality Matches Modality: Pretraining
  Modality-Disentangled Item Representations for Recommendation}. In
  \bibinfo{booktitle}{\emph{Proc. of WWW}}.
\newblock


\bibitem[He et~al\mbox{.}(2016)]%
        {he2016deep}
\bibfield{author}{\bibinfo{person}{Kaiming He}, \bibinfo{person}{Xiangyu
  Zhang}, \bibinfo{person}{Shaoqing Ren}, {and} \bibinfo{person}{Jian Sun}.}
  \bibinfo{year}{2016}\natexlab{}.
\newblock \showarticletitle{Deep residual learning for image recognition}. In
  \bibinfo{booktitle}{\emph{Proc. of CVPR}}.
\newblock


\bibitem[Hou et~al\mbox{.}(2019)]%
        {hou2019explainable}
\bibfield{author}{\bibinfo{person}{Min Hou}, \bibinfo{person}{Le Wu},
  \bibinfo{person}{Enhong Chen}, \bibinfo{person}{Zhi Li},
  \bibinfo{person}{Vincent~W Zheng}, {and} \bibinfo{person}{Qi Liu}.}
  \bibinfo{year}{2019}\natexlab{}.
\newblock \showarticletitle{Explainable fashion recommendation: A semantic
  attribute region guided approach}.
\newblock \bibinfo{journal}{\emph{arXiv preprint arXiv:1905.12862}}
  (\bibinfo{year}{2019}).
\newblock


\bibitem[Hu et~al\mbox{.}(2023)]%
        {hu2023adaptive}
\bibfield{author}{\bibinfo{person}{Hengchang Hu}, \bibinfo{person}{Wei Guo},
  \bibinfo{person}{Yong Liu}, {and} \bibinfo{person}{Min-Yen Kan}.}
  \bibinfo{year}{2023}\natexlab{}.
\newblock \showarticletitle{Adaptive multi-modalities fusion in sequential
  recommendation systems}. In \bibinfo{booktitle}{\emph{Proc. of CIKM}}.
\newblock


\bibitem[Javed et~al\mbox{.}(2021)]%
        {javed2021review}
\bibfield{author}{\bibinfo{person}{Umair Javed}, \bibinfo{person}{Kamran
  Shaukat}, \bibinfo{person}{Ibrahim~A Hameed}, \bibinfo{person}{Farhat Iqbal},
  \bibinfo{person}{Talha~Mahboob Alam}, {and} \bibinfo{person}{Suhuai Luo}.}
  \bibinfo{year}{2021}\natexlab{}.
\newblock \showarticletitle{A review of content-based and context-based
  recommendation systems}.
\newblock \bibinfo{journal}{\emph{International Journal of Emerging
  Technologies in Learning (iJET)}} (\bibinfo{year}{2021}).
\newblock


\bibitem[Jia et~al\mbox{.}(2022)]%
        {jia2022preference}
\bibfield{author}{\bibinfo{person}{Xiangen Jia}, \bibinfo{person}{Yihong Dong},
  \bibinfo{person}{Feng Zhu}, \bibinfo{person}{Yu Xin}, {and}
  \bibinfo{person}{Jiangbo Qian}.} \bibinfo{year}{2022}\natexlab{}.
\newblock \showarticletitle{Preference-corrected multimodal graph convolutional
  recommendation network}.
\newblock \bibinfo{journal}{\emph{Applied Intelligence}}
  (\bibinfo{year}{2022}).
\newblock


\bibitem[Kim et~al\mbox{.}(2022)]%
        {kim2022mario}
\bibfield{author}{\bibinfo{person}{Taeri Kim}, \bibinfo{person}{Yeon-Chang
  Lee}, \bibinfo{person}{Kijung Shin}, {and} \bibinfo{person}{Sang-Wook Kim}.}
  \bibinfo{year}{2022}\natexlab{}.
\newblock \showarticletitle{MARIO: Modality-Aware Attention and
  Modality-Preserving Decoders for Multimedia Recommendation}. In
  \bibinfo{booktitle}{\emph{Proc. of CIKM}}.
\newblock


\bibitem[Kim(2014)]%
        {kim-2014-convolutional}
\bibfield{author}{\bibinfo{person}{Yoon Kim}.} \bibinfo{year}{2014}\natexlab{}.
\newblock \showarticletitle{Convolutional Neural Networks for Sentence
  Classification}. In \bibinfo{booktitle}{\emph{Proc. of EMNLP}}.
  \bibinfo{pages}{1746--1751}.
\newblock
\urldef\tempurl%
\url{https://doi.org/10.3115/v1/D14-1181}
\showDOI{\tempurl}


\bibitem[Koren et~al\mbox{.}(2009)]%
        {koren2009matrix}
\bibfield{author}{\bibinfo{person}{Yehuda Koren}, \bibinfo{person}{Robert
  Bell}, {and} \bibinfo{person}{Chris Volinsky}.}
  \bibinfo{year}{2009}\natexlab{}.
\newblock \showarticletitle{Matrix factorization techniques for recommender
  systems}.
\newblock \bibinfo{journal}{\emph{Computer}} (\bibinfo{year}{2009}).
\newblock


\bibitem[Krishnan et~al\mbox{.}(2020)]%
        {krishnan2020transfer}
\bibfield{author}{\bibinfo{person}{Adit Krishnan}, \bibinfo{person}{Mahashweta
  Das}, \bibinfo{person}{Mangesh Bendre}, \bibinfo{person}{Hao Yang}, {and}
  \bibinfo{person}{Hari Sundaram}.} \bibinfo{year}{2020}\natexlab{}.
\newblock \showarticletitle{Transfer learning via contextual invariants for
  one-to-many cross-domain recommendation}. In \bibinfo{booktitle}{\emph{Proc.
  of SIGIR}}.
\newblock


\bibitem[Lei et~al\mbox{.}(2021)]%
        {lei2021understanding}
\bibfield{author}{\bibinfo{person}{Chenyi Lei}, \bibinfo{person}{Shixian Luo},
  \bibinfo{person}{Yong Liu}, \bibinfo{person}{Wanggui He},
  \bibinfo{person}{Jiamang Wang}, \bibinfo{person}{Guoxin Wang},
  \bibinfo{person}{Haihong Tang}, \bibinfo{person}{Chunyan Miao}, {and}
  \bibinfo{person}{Houqiang Li}.} \bibinfo{year}{2021}\natexlab{}.
\newblock \showarticletitle{Understanding chinese video and language via
  contrastive multimodal pre-training}. In \bibinfo{booktitle}{\emph{Proc. of
  ACM MM}}.
\newblock


\bibitem[Li et~al\mbox{.}(2022)]%
        {li2022miner}
\bibfield{author}{\bibinfo{person}{Jian Li}, \bibinfo{person}{Jieming Zhu},
  \bibinfo{person}{Qiwei Bi}, \bibinfo{person}{Guohao Cai},
  \bibinfo{person}{Lifeng Shang}, \bibinfo{person}{Zhenhua Dong},
  \bibinfo{person}{Xin Jiang}, {and} \bibinfo{person}{Qun Liu}.}
  \bibinfo{year}{2022}\natexlab{}.
\newblock \showarticletitle{MINER: Multi-interest matching network for news
  recommendation}. In \bibinfo{booktitle}{\emph{Proc. of ACL Findings}}.
  \bibinfo{pages}{343--352}.
\newblock


\bibitem[Li et~al\mbox{.}(2023a)]%
        {li2023multimodal}
\bibfield{author}{\bibinfo{person}{Shuaiyang Li}, \bibinfo{person}{Dan Guo},
  \bibinfo{person}{Kang Liu}, \bibinfo{person}{Richang Hong}, {and}
  \bibinfo{person}{Feng Xue}.} \bibinfo{year}{2023}\natexlab{a}.
\newblock \showarticletitle{Multimodal Counterfactual Learning Network for
  Multimedia-based Recommendation}. In \bibinfo{booktitle}{\emph{Proc. of
  SIGIR}}.
\newblock


\bibitem[Li et~al\mbox{.}(2023b)]%
        {li2023imf}
\bibfield{author}{\bibinfo{person}{Xinhang Li}, \bibinfo{person}{Xiangyu Zhao},
  \bibinfo{person}{Jiaxing Xu}, \bibinfo{person}{Yong Zhang}, {and}
  \bibinfo{person}{Chunxiao Xing}.} \bibinfo{year}{2023}\natexlab{b}.
\newblock \showarticletitle{IMF: interactive multimodal fusion model for link
  prediction}. In \bibinfo{booktitle}{\emph{Proc. of WWW}}.
\newblock


\bibitem[Lian et~al\mbox{.}(2021)]%
        {lian2021multi}
\bibfield{author}{\bibinfo{person}{Jianxun Lian}, \bibinfo{person}{Iyad Batal},
  \bibinfo{person}{Zheng Liu}, \bibinfo{person}{Akshay Soni},
  \bibinfo{person}{Eun~Yong Kang}, \bibinfo{person}{Yajun Wang}, {and}
  \bibinfo{person}{Xing Xie}.} \bibinfo{year}{2021}\natexlab{}.
\newblock \showarticletitle{Multi-Interest-Aware User Modeling for Large-Scale
  Sequential Recommendations}.
\newblock \bibinfo{journal}{\emph{arXiv preprint arXiv:2102.09211}}
  (\bibinfo{year}{2021}).
\newblock


\bibitem[Liang et~al\mbox{.}(2023)]%
        {liang2023mmmlp}
\bibfield{author}{\bibinfo{person}{Jiahao Liang}, \bibinfo{person}{Xiangyu
  Zhao}, \bibinfo{person}{Muyang Li}, \bibinfo{person}{Zijian Zhang},
  \bibinfo{person}{Wanyu Wang}, \bibinfo{person}{Haochen Liu}, {and}
  \bibinfo{person}{Zitao Liu}.} \bibinfo{year}{2023}\natexlab{}.
\newblock \showarticletitle{MMMLP: multi-modal multilayer perceptron for
  sequential recommendations}. In \bibinfo{booktitle}{\emph{Proc. of WWW}}.
\newblock


\bibitem[Lin et~al\mbox{.}(2019)]%
        {lin2019explainable}
\bibfield{author}{\bibinfo{person}{Yujie Lin}, \bibinfo{person}{Pengjie Ren},
  \bibinfo{person}{Zhumin Chen}, \bibinfo{person}{Zhaochun Ren},
  \bibinfo{person}{Jun Ma}, {and} \bibinfo{person}{Maarten De~Rijke}.}
  \bibinfo{year}{2019}\natexlab{}.
\newblock \showarticletitle{Explainable outfit recommendation with joint outfit
  matching and comment generation}.
\newblock \bibinfo{journal}{\emph{IEEE Transactions on Knowledge and Data
  Engineering}} (\bibinfo{year}{2019}).
\newblock


\bibitem[Liu(2022)]%
        {liu2022implicit}
\bibfield{author}{\bibinfo{person}{Bo Liu}.} \bibinfo{year}{2022}\natexlab{}.
\newblock \showarticletitle{Implicit semantic-based personalized micro-videos
  recommendation}.
\newblock \bibinfo{journal}{\emph{arXiv preprint arXiv:2205.03297}}
  (\bibinfo{year}{2022}).
\newblock


\bibitem[Liu et~al\mbox{.}(2021a)]%
        {liu2021noninvasive}
\bibfield{author}{\bibinfo{person}{Chang Liu}, \bibinfo{person}{Xiaoguang Li},
  \bibinfo{person}{Guohao Cai}, \bibinfo{person}{Zhenhua Dong},
  \bibinfo{person}{Hong Zhu}, {and} \bibinfo{person}{Lifeng Shang}.}
  \bibinfo{year}{2021}\natexlab{a}.
\newblock \showarticletitle{Noninvasive self-attention for side information
  fusion in sequential recommendation}. In \bibinfo{booktitle}{\emph{Proc. of
  AAAI}}.
\newblock


\bibitem[Liu et~al\mbox{.}(2023)]%
        {liu2023semantic}
\bibfield{author}{\bibinfo{person}{Fan Liu}, \bibinfo{person}{Huilin Chen},
  \bibinfo{person}{Zhiyong Cheng}, \bibinfo{person}{Liqiang Nie}, {and}
  \bibinfo{person}{Mohan Kankanhalli}.} \bibinfo{year}{2023}\natexlab{}.
\newblock \showarticletitle{Semantic-Guided Feature Distillation for Multimodal
  Recommendation}. In \bibinfo{booktitle}{\emph{Proc. of ACM MM}}.
\newblock


\bibitem[Liu et~al\mbox{.}(2022a)]%
        {liu2022disentangled}
\bibfield{author}{\bibinfo{person}{Fan Liu}, \bibinfo{person}{Zhiyong Cheng},
  \bibinfo{person}{Huilin Chen}, \bibinfo{person}{Anan Liu},
  \bibinfo{person}{Liqiang Nie}, {and} \bibinfo{person}{Mohan Kankanhalli}.}
  \bibinfo{year}{2022}\natexlab{a}.
\newblock \showarticletitle{Disentangled Multimodal Representation Learning for
  Recommendation}.
\newblock \bibinfo{journal}{\emph{arXiv preprint arXiv:2203.05406}}
  (\bibinfo{year}{2022}).
\newblock


\bibitem[Liu et~al\mbox{.}(2022b)]%
        {liu2022multi}
\bibfield{author}{\bibinfo{person}{Huizhi Liu}, \bibinfo{person}{Chen Li},
  {and} \bibinfo{person}{Lihua Tian}.} \bibinfo{year}{2022}\natexlab{b}.
\newblock \showarticletitle{Multi-modal Graph Attention Network for Video
  Recommendation}. In \bibinfo{booktitle}{\emph{Proc. of CCET}}.
\newblock


\bibitem[Liu et~al\mbox{.}(2019b)]%
        {liu2019nrpa}
\bibfield{author}{\bibinfo{person}{Hongtao Liu}, \bibinfo{person}{Fangzhao Wu},
  \bibinfo{person}{Wenjun Wang}, \bibinfo{person}{Xianchen Wang},
  \bibinfo{person}{Pengfei Jiao}, \bibinfo{person}{Chuhan Wu}, {and}
  \bibinfo{person}{Xing Xie}.} \bibinfo{year}{2019}\natexlab{b}.
\newblock \showarticletitle{NRPA: neural recommendation with personalized
  attention}. In \bibinfo{booktitle}{\emph{Proc. of SIGIR}}.
\newblock


\bibitem[Liu et~al\mbox{.}(2022d)]%
        {liu2022megcf}
\bibfield{author}{\bibinfo{person}{Kang Liu}, \bibinfo{person}{Feng Xue},
  \bibinfo{person}{Dan Guo}, \bibinfo{person}{Le Wu}, \bibinfo{person}{Shujie
  Li}, {and} \bibinfo{person}{Richang Hong}.} \bibinfo{year}{2022}\natexlab{d}.
\newblock \showarticletitle{MEGCF: Multimodal entity graph collaborative
  filtering for personalized recommendation}.
\newblock \bibinfo{journal}{\emph{TOIS}} (\bibinfo{year}{2022}).
\newblock


\bibitem[Liu et~al\mbox{.}(2019a)]%
        {liu2019user}
\bibfield{author}{\bibinfo{person}{Shang Liu}, \bibinfo{person}{Zhenzhong
  Chen}, \bibinfo{person}{Hongyi Liu}, {and} \bibinfo{person}{Xinghai Hu}.}
  \bibinfo{year}{2019}\natexlab{a}.
\newblock \showarticletitle{User-video co-attention network for personalized
  micro-video recommendation}. In \bibinfo{booktitle}{\emph{Proc. of WWW}}.
\newblock


\bibitem[Liu et~al\mbox{.}(2021b)]%
        {liu2021pre}
\bibfield{author}{\bibinfo{person}{Yong Liu}, \bibinfo{person}{Susen Yang},
  \bibinfo{person}{Chenyi Lei}, \bibinfo{person}{Guoxin Wang},
  \bibinfo{person}{Haihong Tang}, \bibinfo{person}{Juyong Zhang},
  \bibinfo{person}{Aixin Sun}, {and} \bibinfo{person}{Chunyan Miao}.}
  \bibinfo{year}{2021}\natexlab{b}.
\newblock \showarticletitle{Pre-training graph transformer with multimodal side
  information for recommendation}. In \bibinfo{booktitle}{\emph{Proc. of ACM
  MM}}.
\newblock


\bibitem[Liu et~al\mbox{.}(2022c)]%
        {liu2022contrastive}
\bibfield{author}{\bibinfo{person}{Zhuang Liu}, \bibinfo{person}{Yunpu Ma},
  \bibinfo{person}{Matthias Schubert}, \bibinfo{person}{Yuanxin Ouyang}, {and}
  \bibinfo{person}{Zhang Xiong}.} \bibinfo{year}{2022}\natexlab{c}.
\newblock \showarticletitle{Multi-Modal Contrastive Pre-training for
  Recommendation}. In \bibinfo{booktitle}{\emph{Proc. of ICMR}}.
\newblock


\bibitem[Lv et~al\mbox{.}(2019)]%
        {lv2019interest}
\bibfield{author}{\bibinfo{person}{Junmei Lv}, \bibinfo{person}{Bin Song},
  \bibinfo{person}{Jie Guo}, \bibinfo{person}{Xiaojiang Du}, {and}
  \bibinfo{person}{Mohsen Guizani}.} \bibinfo{year}{2019}\natexlab{}.
\newblock \showarticletitle{Interest-related item similarity model based on
  multimodal data for top-N recommendation}.
\newblock \bibinfo{journal}{\emph{IEEE Access}} (\bibinfo{year}{2019}).
\newblock


\bibitem[Ma et~al\mbox{.}(2019)]%
        {ma2019learning}
\bibfield{author}{\bibinfo{person}{Jianxin Ma}, \bibinfo{person}{Chang Zhou},
  \bibinfo{person}{Peng Cui}, \bibinfo{person}{Hongxia Yang}, {and}
  \bibinfo{person}{Wenwu Zhu}.} \bibinfo{year}{2019}\natexlab{}.
\newblock \showarticletitle{Learning disentangled representations for
  recommendation}.
\newblock \bibinfo{journal}{\emph{Proc. of NeurIPS}} (\bibinfo{year}{2019}).
\newblock


\bibitem[Ma et~al\mbox{.}(2022)]%
        {ma2022crosscbr}
\bibfield{author}{\bibinfo{person}{Yunshan Ma}, \bibinfo{person}{Yingzhi He},
  \bibinfo{person}{An Zhang}, \bibinfo{person}{Xiang Wang}, {and}
  \bibinfo{person}{Tat-Seng Chua}.} \bibinfo{year}{2022}\natexlab{}.
\newblock \showarticletitle{CrossCBR: Cross-view Contrastive Learning for
  Bundle Recommendation}.
\newblock \bibinfo{journal}{\emph{arXiv preprint arXiv:2206.00242}}
  (\bibinfo{year}{2022}).
\newblock


\bibitem[Mikolov et~al\mbox{.}(2013)]%
        {mikolov2013efficient}
\bibfield{author}{\bibinfo{person}{Tomas Mikolov}, \bibinfo{person}{Kai Chen},
  \bibinfo{person}{Greg Corrado}, {and} \bibinfo{person}{Jeffrey Dean}.}
  \bibinfo{year}{2013}\natexlab{}.
\newblock \showarticletitle{Efficient estimation of word representations in
  vector space}.
\newblock \bibinfo{journal}{\emph{arXiv preprint arXiv:1301.3781}}
  (\bibinfo{year}{2013}).
\newblock


\bibitem[Mu et~al\mbox{.}(2022)]%
        {mu2022learning}
\bibfield{author}{\bibinfo{person}{Zongshen Mu}, \bibinfo{person}{Yueting
  Zhuang}, \bibinfo{person}{Jie Tan}, \bibinfo{person}{Jun Xiao}, {and}
  \bibinfo{person}{Siliang Tang}.} \bibinfo{year}{2022}\natexlab{}.
\newblock \showarticletitle{Learning Hybrid Behavior Patterns for Multimedia
  Recommendation}. In \bibinfo{booktitle}{\emph{Proc. of ACM MM}}.
\newblock


\bibitem[Ni et~al\mbox{.}(2022)]%
        {ni2022two}
\bibfield{author}{\bibinfo{person}{Juan Ni}, \bibinfo{person}{Zhenhua Huang},
  \bibinfo{person}{Yang Hu}, {and} \bibinfo{person}{Chen Lin}.}
  \bibinfo{year}{2022}\natexlab{}.
\newblock \showarticletitle{A two-stage embedding model for recommendation with
  multimodal auxiliary information}.
\newblock \bibinfo{journal}{\emph{Information Sciences}}
  (\bibinfo{year}{2022}).
\newblock


\bibitem[Pan et~al\mbox{.}(2022)]%
        {pan2022multimodal}
\bibfield{author}{\bibinfo{person}{Xingyu Pan}, \bibinfo{person}{Yushuo Chen},
  \bibinfo{person}{Changxin Tian}, \bibinfo{person}{Zihan Lin},
  \bibinfo{person}{Jinpeng Wang}, \bibinfo{person}{He Hu}, {and}
  \bibinfo{person}{Wayne~Xin Zhao}.} \bibinfo{year}{2022}\natexlab{}.
\newblock \showarticletitle{Multimodal Meta-Learning for Cold-Start Sequential
  Recommendation}. In \bibinfo{booktitle}{\emph{Proc. of CIKM}}.
\newblock


\bibitem[Pennington et~al\mbox{.}(2014)]%
        {pennington2014glove}
\bibfield{author}{\bibinfo{person}{Jeffrey Pennington},
  \bibinfo{person}{Richard Socher}, {and} \bibinfo{person}{Christopher~D
  Manning}.} \bibinfo{year}{2014}\natexlab{}.
\newblock \showarticletitle{Glove: Global vectors for word representation}. In
  \bibinfo{booktitle}{\emph{Proc. of EMNLP}}. \bibinfo{pages}{1532--1543}.
\newblock


\bibitem[Radford et~al\mbox{.}(2021)]%
        {radford2021learning}
\bibfield{author}{\bibinfo{person}{Alec Radford}, \bibinfo{person}{Jong~Wook
  Kim}, \bibinfo{person}{Chris Hallacy}, \bibinfo{person}{Aditya Ramesh},
  \bibinfo{person}{Gabriel Goh}, \bibinfo{person}{Sandhini Agarwal},
  \bibinfo{person}{Girish Sastry}, \bibinfo{person}{Amanda Askell},
  \bibinfo{person}{Pamela Mishkin}, \bibinfo{person}{Jack Clark},
  {et~al\mbox{.}}} \bibinfo{year}{2021}\natexlab{}.
\newblock \showarticletitle{Learning transferable visual models from natural
  language supervision}. In \bibinfo{booktitle}{\emph{Proc. of ICML}}.
\newblock


\bibitem[Salah et~al\mbox{.}(2020)]%
        {salah2020cornac}
\bibfield{author}{\bibinfo{person}{Aghiles Salah}, \bibinfo{person}{Quoc-Tuan
  Truong}, {and} \bibinfo{person}{Hady~W Lauw}.}
  \bibinfo{year}{2020}\natexlab{}.
\newblock \showarticletitle{Cornac: A Comparative Framework for Multimodal
  Recommender Systems}.
\newblock \bibinfo{journal}{\emph{Journal of Machine Learning Research}}
  (\bibinfo{year}{2020}).
\newblock


\bibitem[Shang et~al\mbox{.}(2023)]%
        {shang2023enhancing}
\bibfield{author}{\bibinfo{person}{Yu Shang}, \bibinfo{person}{Chen Gao},
  \bibinfo{person}{Jiansheng Chen}, \bibinfo{person}{Depeng Jin},
  \bibinfo{person}{Huimin Ma}, {and} \bibinfo{person}{Yong Li}.}
  \bibinfo{year}{2023}\natexlab{}.
\newblock \showarticletitle{Enhancing Adversarial Robustness of Multi-modal
  Recommendation via Modality Balancing}. In \bibinfo{booktitle}{\emph{Proc. of
  ACM MM}}.
\newblock


\bibitem[Sun et~al\mbox{.}(2020)]%
        {sun2020multi}
\bibfield{author}{\bibinfo{person}{Rui Sun}, \bibinfo{person}{Xuezhi Cao},
  \bibinfo{person}{Yan Zhao}, \bibinfo{person}{Junchen Wan},
  \bibinfo{person}{Kun Zhou}, \bibinfo{person}{Fuzheng Zhang},
  \bibinfo{person}{Zhongyuan Wang}, {and} \bibinfo{person}{Kai Zheng}.}
  \bibinfo{year}{2020}\natexlab{}.
\newblock \showarticletitle{Multi-modal knowledge graphs for recommender
  systems}. In \bibinfo{booktitle}{\emph{Proc. of CIKM}}.
\newblock


\bibitem[Tao et~al\mbox{.}(2020)]%
        {tao2020mgat}
\bibfield{author}{\bibinfo{person}{Zhulin Tao}, \bibinfo{person}{Yinwei Wei},
  \bibinfo{person}{Xiang Wang}, \bibinfo{person}{Xiangnan He},
  \bibinfo{person}{Xianglin Huang}, {and} \bibinfo{person}{Tat-Seng Chua}.}
  \bibinfo{year}{2020}\natexlab{}.
\newblock \showarticletitle{MGAT: multimodal graph attention network for
  recommendation}.
\newblock \bibinfo{journal}{\emph{Information Processing \& Management}}
  (\bibinfo{year}{2020}).
\newblock


\bibitem[Vaswani et~al\mbox{.}(2017)]%
        {vaswani2017attention}
\bibfield{author}{\bibinfo{person}{Ashish Vaswani}, \bibinfo{person}{Noam
  Shazeer}, \bibinfo{person}{Niki Parmar}, \bibinfo{person}{Jakob Uszkoreit},
  \bibinfo{person}{Llion Jones}, \bibinfo{person}{Aidan~N Gomez},
  \bibinfo{person}{{\L}ukasz Kaiser}, {and} \bibinfo{person}{Illia
  Polosukhin}.} \bibinfo{year}{2017}\natexlab{}.
\newblock \showarticletitle{Attention is all you need}.
\newblock \bibinfo{journal}{\emph{Proc. of NeurIPS}} (\bibinfo{year}{2017}).
\newblock


\bibitem[Wang et~al\mbox{.}(2019)]%
        {wang2019multi}
\bibfield{author}{\bibinfo{person}{Hongwei Wang}, \bibinfo{person}{Fuzheng
  Zhang}, \bibinfo{person}{Miao Zhao}, \bibinfo{person}{Wenjie Li},
  \bibinfo{person}{Xing Xie}, {and} \bibinfo{person}{Minyi Guo}.}
  \bibinfo{year}{2019}\natexlab{}.
\newblock \showarticletitle{Multi-task feature learning for knowledge graph
  enhanced recommendation}. In \bibinfo{booktitle}{\emph{Proc. of WWW}}.
\newblock


\bibitem[Wang et~al\mbox{.}(2022b)]%
        {wang2022transrec}
\bibfield{author}{\bibinfo{person}{Jie Wang}, \bibinfo{person}{Fajie Yuan},
  \bibinfo{person}{Mingyue Cheng}, \bibinfo{person}{Joemon~M Jose},
  \bibinfo{person}{Chenyun Yu}, \bibinfo{person}{Beibei Kong},
  \bibinfo{person}{Zhijin Wang}, \bibinfo{person}{Bo Hu}, {and}
  \bibinfo{person}{Zang Li}.} \bibinfo{year}{2022}\natexlab{b}.
\newblock \showarticletitle{TransRec: Learning Transferable Recommendation from
  Mixture-of-Modality Feedback}.
\newblock \bibinfo{journal}{\emph{arXiv preprint arXiv:2206.06190}}
  (\bibinfo{year}{2022}).
\newblock


\bibitem[Wang et~al\mbox{.}(2023b)]%
        {wang2023large}
\bibfield{author}{\bibinfo{person}{Maolin Wang}, \bibinfo{person}{Yao Zhao},
  \bibinfo{person}{Jiajia Liu}, \bibinfo{person}{Jingdong Chen},
  \bibinfo{person}{Chenyi Zhuang}, \bibinfo{person}{Jinjie Gu},
  \bibinfo{person}{Ruocheng Guo}, {and} \bibinfo{person}{Xiangyu Zhao}.}
  \bibinfo{year}{2023}\natexlab{b}.
\newblock \showarticletitle{Large Multimodal Model Compression via Efficient
  Pruning and Distillation at AntGroup}.
\newblock \bibinfo{journal}{\emph{arXiv preprint arXiv:2312.05795}}
  (\bibinfo{year}{2023}).
\newblock


\bibitem[Wang et~al\mbox{.}(2022a)]%
        {wang2022multimodal}
\bibfield{author}{\bibinfo{person}{Peng Wang}, \bibinfo{person}{Jiangheng Wu},
  {and} \bibinfo{person}{Xiaohang Chen}.} \bibinfo{year}{2022}\natexlab{a}.
\newblock \showarticletitle{Multimodal Entity Linking with Gated Hierarchical
  Fusion and Contrastive Training}. In \bibinfo{booktitle}{\emph{Proc. of
  SIGIR}}.
\newblock


\bibitem[Wang et~al\mbox{.}(2021b)]%
        {wang2021dualgnn}
\bibfield{author}{\bibinfo{person}{Qifan Wang}, \bibinfo{person}{Yinwei Wei},
  \bibinfo{person}{Jianhua Yin}, \bibinfo{person}{Jianlong Wu},
  \bibinfo{person}{Xuemeng Song}, {and} \bibinfo{person}{Liqiang Nie}.}
  \bibinfo{year}{2021}\natexlab{b}.
\newblock \showarticletitle{Dualgnn: Dual graph neural network for multimedia
  recommendation}.
\newblock \bibinfo{journal}{\emph{IEEE Transactions on Multimedia}}
  (\bibinfo{year}{2021}).
\newblock


\bibitem[Wang et~al\mbox{.}(2023a)]%
        {9720218}
\bibfield{author}{\bibinfo{person}{Xin Wang}, \bibinfo{person}{Hong Chen},
  \bibinfo{person}{Yuwei Zhou}, \bibinfo{person}{Jianxin Ma}, {and}
  \bibinfo{person}{Wenwu Zhu}.} \bibinfo{year}{2023}\natexlab{a}.
\newblock \showarticletitle{Disentangled Representation Learning for
  Recommendation}.
\newblock \bibinfo{journal}{\emph{IEEE Transactions on Pattern Analysis and
  Machine Intelligence}} (\bibinfo{year}{2023}).
\newblock


\bibitem[Wang et~al\mbox{.}(2021a)]%
        {wang2021multimodal}
\bibfield{author}{\bibinfo{person}{Xin Wang}, \bibinfo{person}{Hong Chen},
  {and} \bibinfo{person}{Wenwu Zhu}.} \bibinfo{year}{2021}\natexlab{a}.
\newblock \showarticletitle{Multimodal disentangled representation for
  recommendation}. In \bibinfo{booktitle}{\emph{Proc. of ICME}}.
\newblock


\bibitem[Wang et~al\mbox{.}(2020)]%
        {wang2020enhanced}
\bibfield{author}{\bibinfo{person}{Yuequn Wang}, \bibinfo{person}{Liyan Dong},
  \bibinfo{person}{Hao Zhang}, \bibinfo{person}{Xintao Ma},
  \bibinfo{person}{Yongli Li}, {and} \bibinfo{person}{Minghui Sun}.}
  \bibinfo{year}{2020}\natexlab{}.
\newblock \showarticletitle{An enhanced multi-modal recommendation based on
  alternate training with knowledge graph representation}.
\newblock \bibinfo{journal}{\emph{Ieee Access}} (\bibinfo{year}{2020}).
\newblock


\bibitem[Wei et~al\mbox{.}(2023)]%
        {wei2023multi}
\bibfield{author}{\bibinfo{person}{Wei Wei}, \bibinfo{person}{Chao Huang},
  \bibinfo{person}{Lianghao Xia}, {and} \bibinfo{person}{Chuxu Zhang}.}
  \bibinfo{year}{2023}\natexlab{}.
\newblock \showarticletitle{Multi-modal self-supervised learning for
  recommendation}. In \bibinfo{booktitle}{\emph{Proc. of WWW}}.
\newblock


\bibitem[Wei et~al\mbox{.}(2024)]%
        {wei2024promptmm}
\bibfield{author}{\bibinfo{person}{Wei Wei}, \bibinfo{person}{Jiabin Tang},
  \bibinfo{person}{Lianghao Xia}, \bibinfo{person}{Yangqin Jiang}, {and}
  \bibinfo{person}{Chao Huang}.} \bibinfo{year}{2024}\natexlab{}.
\newblock \showarticletitle{PromptMM: Multi-Modal Knowledge Distillation for
  Recommendation with Prompt-Tuning}. In \bibinfo{booktitle}{\emph{Proc. of
  WWW}}.
\newblock


\bibitem[Wei et~al\mbox{.}(2019)]%
        {wei2019mmgcn}
\bibfield{author}{\bibinfo{person}{Yinwei Wei}, \bibinfo{person}{Xiang Wang},
  \bibinfo{person}{Liqiang Nie}, \bibinfo{person}{Xiangnan He},
  \bibinfo{person}{Richang Hong}, {and} \bibinfo{person}{Tat-Seng Chua}.}
  \bibinfo{year}{2019}\natexlab{}.
\newblock \showarticletitle{MMGCN: Multi-modal graph convolution network for
  personalized recommendation of micro-video}. In
  \bibinfo{booktitle}{\emph{Proc. of ACM MM}}.
\newblock


\bibitem[Wu et~al\mbox{.}(2021)]%
        {wu2021mm}
\bibfield{author}{\bibinfo{person}{Chuhan Wu}, \bibinfo{person}{Fangzhao Wu},
  \bibinfo{person}{Tao Qi}, {and} \bibinfo{person}{Yongfeng Huang}.}
  \bibinfo{year}{2021}\natexlab{}.
\newblock \showarticletitle{Mm-rec: multimodal news recommendation}.
\newblock \bibinfo{journal}{\emph{arXiv preprint arXiv:2104.07407}}
  (\bibinfo{year}{2021}).
\newblock


\bibitem[Xiao et~al\mbox{.}(2020)]%
        {xiao2020deep}
\bibfield{author}{\bibinfo{person}{Zhibo Xiao}, \bibinfo{person}{Luwei Yang},
  \bibinfo{person}{Wen Jiang}, \bibinfo{person}{Yi Wei}, \bibinfo{person}{Yi
  Hu}, {and} \bibinfo{person}{Hao Wang}.} \bibinfo{year}{2020}\natexlab{}.
\newblock \showarticletitle{Deep multi-interest network for click-through rate
  prediction}. In \bibinfo{booktitle}{\emph{Proc. of CIKM}}.
  \bibinfo{pages}{2265--2268}.
\newblock


\bibitem[Xu et~al\mbox{.}(2020)]%
        {xu2020recommendation}
\bibfield{author}{\bibinfo{person}{Cai Xu}, \bibinfo{person}{Ziyu Guan},
  \bibinfo{person}{Wei Zhao}, \bibinfo{person}{Quanzhou Wu},
  \bibinfo{person}{Meng Yan}, \bibinfo{person}{Long Chen}, {and}
  \bibinfo{person}{Qiguang Miao}.} \bibinfo{year}{2020}\natexlab{}.
\newblock \showarticletitle{Recommendation by users’ multimodal preferences
  for smart city applications}.
\newblock \bibinfo{journal}{\emph{IEEE Transactions on Industrial Informatics}}
  (\bibinfo{year}{2020}).
\newblock


\bibitem[Yi and Chen(2021)]%
        {yi2021multi}
\bibfield{author}{\bibinfo{person}{Jing Yi} {and} \bibinfo{person}{Zhenzhong
  Chen}.} \bibinfo{year}{2021}\natexlab{}.
\newblock \showarticletitle{Multi-Modal Variational Graph Auto-Encoder for
  Recommendation Systems}.
\newblock \bibinfo{journal}{\emph{IEEE Transactions on Multimedia}}
  (\bibinfo{year}{2021}).
\newblock


\bibitem[Yi et~al\mbox{.}(2022)]%
        {yi2022multi}
\bibfield{author}{\bibinfo{person}{Zixuan Yi}, \bibinfo{person}{Xi Wang},
  \bibinfo{person}{Iadh Ounis}, {and} \bibinfo{person}{Craig Macdonald}.}
  \bibinfo{year}{2022}\natexlab{}.
\newblock \showarticletitle{Multi-modal graph contrastive learning for
  micro-video recommendation}. In \bibinfo{booktitle}{\emph{Proc. of SIGIR}}.
\newblock


\bibitem[Yinwei et~al\mbox{.}(2021)]%
        {yinwei2021grcn}
\bibfield{author}{\bibinfo{person}{Wei Yinwei}, \bibinfo{person}{Wang Xiang},
  \bibinfo{person}{Nie Liqiang}, \bibinfo{person}{He Xiangnan}, {and}
  \bibinfo{person}{Chua Tat-Seng}.} \bibinfo{year}{2021}\natexlab{}.
\newblock \showarticletitle{GRCN: Graph-Refined Convolutional Network for
  Multimedia Recommendation with Implicit Feedback}.
\newblock \bibinfo{journal}{\emph{arXiv preprint arXiv:2111.02036}}
  (\bibinfo{year}{2021}).
\newblock


\bibitem[Yu et~al\mbox{.}(2023)]%
        {yu2023multi}
\bibfield{author}{\bibinfo{person}{Penghang Yu}, \bibinfo{person}{Zhiyi Tan},
  \bibinfo{person}{Guanming Lu}, {and} \bibinfo{person}{Bing-Kun Bao}.}
  \bibinfo{year}{2023}\natexlab{}.
\newblock \showarticletitle{Multi-view graph convolutional network for
  multimedia recommendation}. In \bibinfo{booktitle}{\emph{Proc. of ACM MM}}.
\newblock


\bibitem[Zhang et~al\mbox{.}(2021)]%
        {zhang2021mining}
\bibfield{author}{\bibinfo{person}{Jinghao Zhang}, \bibinfo{person}{Yanqiao
  Zhu}, \bibinfo{person}{Qiang Liu}, \bibinfo{person}{Shu Wu},
  \bibinfo{person}{Shuhui Wang}, {and} \bibinfo{person}{Liang Wang}.}
  \bibinfo{year}{2021}\natexlab{}.
\newblock \showarticletitle{Mining latent structures for multimedia
  recommendation}. In \bibinfo{booktitle}{\emph{Proc. of ACM MM}}.
\newblock


\bibitem[Zhang et~al\mbox{.}(2022)]%
        {zhang2022latent}
\bibfield{author}{\bibinfo{person}{Jinghao Zhang}, \bibinfo{person}{Yanqiao
  Zhu}, \bibinfo{person}{Qiang Liu}, \bibinfo{person}{Mengqi Zhang},
  \bibinfo{person}{Shu Wu}, {and} \bibinfo{person}{Liang Wang}.}
  \bibinfo{year}{2022}\natexlab{}.
\newblock \showarticletitle{Latent Structure Mining with Contrastive Modality
  Fusion for Multimedia Recommendation}.
\newblock \bibinfo{journal}{\emph{IEEE Transactions on Knowledge and Data
  Engineering}} (\bibinfo{year}{2022}).
\newblock


\bibitem[Zhang et~al\mbox{.}(2023)]%
        {zhang2023multimodal}
\bibfield{author}{\bibinfo{person}{Lingzi Zhang}, \bibinfo{person}{Xin Zhou},
  {and} \bibinfo{person}{Zhiqi Shen}.} \bibinfo{year}{2023}\natexlab{}.
\newblock \showarticletitle{Multimodal Pre-training Framework for Sequential
  Recommendation via Contrastive Learning}.
\newblock \bibinfo{journal}{\emph{arXiv preprint arXiv:2303.11879}}
  (\bibinfo{year}{2023}).
\newblock


\bibitem[Zhang et~al\mbox{.}(2020a)]%
        {zhang2020multi}
\bibfield{author}{\bibinfo{person}{Xiaoyan Zhang}, \bibinfo{person}{Haihua
  Luo}, \bibinfo{person}{Bowei Chen}, {and} \bibinfo{person}{Guibing Guo}.}
  \bibinfo{year}{2020}\natexlab{a}.
\newblock \showarticletitle{Multi-view visual Bayesian personalized ranking for
  restaurant recommendation}.
\newblock \bibinfo{journal}{\emph{Applied Intelligence}}
  (\bibinfo{year}{2020}).
\newblock


\bibitem[Zhang et~al\mbox{.}(2020b)]%
        {zhang2020content}
\bibfield{author}{\bibinfo{person}{Yin Zhang}, \bibinfo{person}{Ziwei Zhu},
  \bibinfo{person}{Yun He}, {and} \bibinfo{person}{James Caverlee}.}
  \bibinfo{year}{2020}\natexlab{b}.
\newblock \showarticletitle{Content-collaborative disentanglement
  representation learning for enhanced recommendation}. In
  \bibinfo{booktitle}{\emph{Proc. of RecSys}}.
\newblock


\bibitem[Zhao and Wang(2021)]%
        {zhao2021multimodal}
\bibfield{author}{\bibinfo{person}{Feng Zhao} {and} \bibinfo{person}{Donglin
  Wang}.} \bibinfo{year}{2021}\natexlab{}.
\newblock \showarticletitle{Multimodal graph meta contrastive learning}. In
  \bibinfo{booktitle}{\emph{Proc. of CIKM}}.
\newblock


\bibitem[Zhong et~al\mbox{.}(2024)]%
        {zhong2024mirror}
\bibfield{author}{\bibinfo{person}{Shanshan Zhong}, \bibinfo{person}{Zhongzhan
  Huang}, \bibinfo{person}{Daifeng Li}, \bibinfo{person}{Wushao Wen},
  \bibinfo{person}{Jinghui Qin}, {and} \bibinfo{person}{Liang Lin}.}
  \bibinfo{year}{2024}\natexlab{}.
\newblock \showarticletitle{Mirror Gradient: Towards Robust Multimodal
  Recommender Systems via Exploring Flat Local Minima}. In
  \bibinfo{booktitle}{\emph{Proc. of WWW}}.
\newblock


\bibitem[Zhou et~al\mbox{.}(2023c)]%
        {zhou2023comprehensive}
\bibfield{author}{\bibinfo{person}{Hongyu Zhou}, \bibinfo{person}{Xin Zhou},
  \bibinfo{person}{Zhiwei Zeng}, \bibinfo{person}{Lingzi Zhang}, {and}
  \bibinfo{person}{Zhiqi Shen}.} \bibinfo{year}{2023}\natexlab{c}.
\newblock \showarticletitle{A Comprehensive Survey on Multimodal Recommender
  Systems: Taxonomy, Evaluation, and Future Directions}.
\newblock \bibinfo{journal}{\emph{arXiv preprint arXiv:2302.04473}}
  (\bibinfo{year}{2023}).
\newblock


\bibitem[Zhou(2022)]%
        {zhou2022tale}
\bibfield{author}{\bibinfo{person}{Xin Zhou}.} \bibinfo{year}{2022}\natexlab{}.
\newblock \showarticletitle{A Tale of Two Graphs: Freezing and Denoising Graph
  Structures for Multimodal Recommendation}.
\newblock \bibinfo{journal}{\emph{arXiv preprint arXiv:2211.06924}}
  (\bibinfo{year}{2022}).
\newblock


\bibitem[Zhou(2023)]%
        {zhou2023mmrec}
\bibfield{author}{\bibinfo{person}{Xin Zhou}.} \bibinfo{year}{2023}\natexlab{}.
\newblock \showarticletitle{MMRec: Simplifying Multimodal Recommendation}.
\newblock \bibinfo{journal}{\emph{arXiv preprint arXiv:2302.03497}}
  (\bibinfo{year}{2023}).
\newblock


\bibitem[Zhou et~al\mbox{.}(2023b)]%
        {zhou2023bootstrap}
\bibfield{author}{\bibinfo{person}{Xin Zhou}, \bibinfo{person}{Hongyu Zhou},
  \bibinfo{person}{Yong Liu}, \bibinfo{person}{Zhiwei Zeng},
  \bibinfo{person}{Chunyan Miao}, \bibinfo{person}{Pengwei Wang},
  \bibinfo{person}{Yuan You}, {and} \bibinfo{person}{Feijun Jiang}.}
  \bibinfo{year}{2023}\natexlab{b}.
\newblock \showarticletitle{Bootstrap latent representations for multi-modal
  recommendation}. In \bibinfo{booktitle}{\emph{Proc. of WWW}}.
\newblock


\bibitem[Zhou et~al\mbox{.}(2023a)]%
        {zhou2023attention}
\bibfield{author}{\bibinfo{person}{Yan Zhou}, \bibinfo{person}{Jie Guo},
  \bibinfo{person}{Hao Sun}, \bibinfo{person}{Bin Song}, {and}
  \bibinfo{person}{Fei~Richard Yu}.} \bibinfo{year}{2023}\natexlab{a}.
\newblock \showarticletitle{Attention-guided multi-step fusion: a hierarchical
  fusion network for multimodal recommendation}. In
  \bibinfo{booktitle}{\emph{Proc. of SIGIR}}.
\newblock


\bibitem[Zhu et~al\mbox{.}(2022)]%
        {zhu2022combo}
\bibfield{author}{\bibinfo{person}{Chenxu Zhu}, \bibinfo{person}{Peng Du},
  \bibinfo{person}{Weinan Zhang}, \bibinfo{person}{Yong Yu}, {and}
  \bibinfo{person}{Yang Cao}.} \bibinfo{year}{2022}\natexlab{}.
\newblock \showarticletitle{Combo-Fashion: Fashion Clothes Matching CTR
  Prediction with Item History}. In \bibinfo{booktitle}{\emph{Proc. of KDD}}.
\newblock


\end{thebibliography}


\end{document}